\documentclass[12pt,a4paper]{article}
\usepackage{amssymb}
\usepackage{amsmath}
\usepackage[margin=2cm,nohead]{geometry}

\allowdisplaybreaks[1]
\input{tcilatex}
\begin{document}

\title{$CP$--Violation in $K,$ $B$ and $B_{s}$ decays}
\author{Fayyazuddin \\
National Centre for Physics and Physics Department Quaid-i-Azam University,\\
Islamabad}
\date{fayyazuddins@gmail.com}
\maketitle

\begin{abstract}
In this review we give an overview of $CP$-violation for $K^{0}(\bar{K}%
^{0}), $ $B_{q}^{0}(\bar{B}_{q}^{0}),$ $q=d,s$ systems. Direct $CP-$%
violation and mixing induced $CP$-violation are discussed.
\end{abstract}

\section{Introduction}

Symmetries have played an important role in particle physics. In quantum
mechanics a symmetry is associated with a group of transformations under
which a Lagrangian remains invariant. Symmetries limit the possible terms in
a Lagrangian and are associated with conservation laws. Here we will be
concerned with the role of discrete symmetries: Space Reflection (Parity) $P$%
: $\vec{x}\rightarrow -\vec{x}$, Time Reversal $T$: $t\rightarrow -t$ and
Charge Conjugation $C$: $particle\rightarrow antiparticle$.

Quantum Electrodynamics (QED) and Quantum Chromodynamics (QCD) respect all
these symmetries. Also, all Lorentz invariant local quantum field theories
are $CPT$ invariant. However, in weak interactions $C$ and $P$ are maximally
violated separately but as we will see below, $CP$ is conserved.

First indication of parity violation was revealed in the decay of a particle
with spin parity $J^{P}=0^{-},$ called $K$-meson into two modes $%
K^{0}\rightarrow \pi ^{+}\pi ^{-}$ (parity violating), and $K^{0}\rightarrow
\pi ^{+}\pi ^{-}$ $\pi ^{0}$(parity conserving).

Lee and Yang in 1956, suggested that there is no experimental evidence for
parity conservation in weak interaction. They suggested number of
experiments to test the validity of space reflection invariance in weak
decays. One way to test this is to measure the helicity of outgoing muon in
the decay: 
\begin{equation*}
\pi ^{+}\rightarrow \mu ^{+}+\nu _{\mu }
\end{equation*}

The helicity of muon comes out to be negative, showing that parity
conservation does not hold in this decay. In the rest frame of the pion,
since $\mu^{+}$ comes out with negative helicity, the neutrino must also
come out with negative helicity because of the spin conservation. Thus
confirming the fact that neutrino is left handed. 
\begin{equation*}
\pi ^{+}\rightarrow \mu ^{+}(-)+\nu _{\mu }
\end{equation*}

Under charge conjugation,

\begin{equation*}
\pi^{+}\overset{C}{\rightarrow}\pi ^{-} \qquad \mu^{+}\overset{C}{\rightarrow%
}\mu ^{-} \qquad \nu_{\mu }\overset{C}{\rightarrow}\bar{\nu}_{\mu}
\end{equation*}

Helicity $\mathcal{H}=\frac{\vec{\sigma}\cdot \vec{p}}{\left| \vec{p}\right| 
}$ under $C$ and $P$ transforms as,

\begin{equation*}
\mathcal{H} \overset{C}{\rightarrow} \mathcal{H}, \qquad \mathcal{H} \overset%
{P}{\rightarrow} -\mathcal{H}
\end{equation*}

Invariance under $C$ gives,

\begin{equation*}
\Gamma _{\pi ^{+}\rightarrow \mu ^{+}(-)\nu _{\mu }}=\Gamma _{\pi
^{-}\rightarrow \mu ^{-}(-)\bar{\nu}_{\mu }}
\end{equation*}

Experimentally,

\begin{equation*}
\Gamma _{\pi ^{+}\rightarrow \mu ^{+}(-)\nu _{\mu }}>>\Gamma _{\pi
^{-}\rightarrow \mu ^{-}(-)\bar{\nu}_{\mu }}
\end{equation*}

showing that $C$ is also violated in weak interactions. However, under $CP$,

\begin{equation*}
\Gamma _{\pi ^{+}\rightarrow \mu ^{+}(-)\nu _{\mu }}\overset{CP}{\rightarrow}
\text{ \ }\Gamma _{\pi ^{-}\rightarrow \mu ^{-}(+)\bar{\nu}_{\mu }}
\end{equation*}

which is seen experimentally. Thus, $CP$ conservation holds in weak
interaction.

In the Standard Model, the fermions for each generation in their left handed
chirality state belong to the representation, 
\begin{eqnarray*}
\left( 
\begin{array}{c}
u_{i} \\ 
d_{i}%
\end{array}%
\right) &:&q(3,2,1/3) \\
\bar{u}_{i} &:&(\bar{3},1,-4/3) \\
\bar{d}_{i} &:&(\bar{3},1,2/3) \\
\left( 
\begin{array}{c}
\nu _{e^{-}} \\ 
e_{i}^{-}%
\end{array}%
\right) &:&l(1,2,-1/2) \\
e_{i}^{+} &:&(1,1,1)
\end{eqnarray*}

of the electroweak unification group $SU_{C}(3)\times SU_{L}(2)\times
U_{Y}(1)$. Hence, the weak interaction Lagrangian for the charged current in
the Standard Model is given by,

\begin{equation*}
\mathcal{L}_{W}=\bar{\psi}_{i}\gamma ^{\mu }(1-\gamma ^{5})\psi _{j}W_{\mu
}^{+}+h.c.
\end{equation*}

where $\psi _{i}$ is any of the left-handed doublet ($i$ is the generation
index). We note that the weak eigenstates $d^{\prime}, s^{\prime}$ and $%
b^{\prime}$ are not equal to the mass eigenstates $d, s$ and $b$. They are
related to each other by a unitarity transformation,

\begin{equation}
\left( 
\begin{array}{c}
d^{\prime } \\ 
s^{\prime } \\ 
b^{\prime }%
\end{array}
\right) =V\left( 
\begin{array}{c}
d \\ 
s \\ 
b%
\end{array}
\right)  \label{05}
\end{equation}

where $V$ is called the $CKM$ matrix.

\begin{equation*}
V=\left( 
\begin{array}{ccc}
V_{ud} & V_{us} & V_{ub} \\ 
V_{cd} & V_{cs} & V_{cb} \\ 
V_{td} & V_{ts} & V_{tb}%
\end{array}
\right)
\end{equation*}

\begin{equation}
\simeq \left( 
\begin{array}{ccc}
1-\frac{1}{2}\lambda ^{2} & \lambda & A\lambda ^{3}\left( \rho -i\eta \right)
\\ 
-\lambda & 1-\frac{1}{2}\lambda ^{2} & A\lambda ^{2} \\ 
A\lambda ^{3}\left( 1-\rho -i\eta \right) & -A\lambda ^{2} & 1%
\end{array}
\right) +O\left( \lambda ^{4}\right) ,\,\left. \lambda =0.22\right.
\label{06}
\end{equation}


The unitarity of $V$, $VV^{\dagger }=1$ gives,$\left[ \text{Fig.1}\right] $

\begin{equation}
V_{ud}^{\ast }V_{ub}+V_{cb}^{\ast }V_{cd}+V_{td}^{\ast }V_{tb}=0  \label{07}
\end{equation}

The second line in equation (\ref{06}) expresses $V$ in terms of Wolfenstien
parametrization. Thus, 
\begin{eqnarray*}
V_{cb} &=&A\lambda ^{2} \\
V_{ub} &=&\left| V_{ub}\right| e^{-i\gamma } \\
V_{td} &=&\left| V_{td}\right| e^{-i\beta }
\end{eqnarray*}

where, 
\begin{equation*}
\tan {\gamma }=\frac{\eta }{\rho }=\frac{\bar{\eta}}{\bar{\rho}},\quad \tan {%
\beta }=\frac{\bar{\eta}}{1-\bar{\rho}},\quad \bar{\rho}=\rho (1-\frac{%
\lambda ^{2}}{2}),\quad \bar{\eta}=\eta (1-\frac{\lambda ^{2}}{2}).
\end{equation*}

In order to show that $\mathcal{L}_{W}$ is $CP$-invariant, we first note
that under $C$, $P$ and $T$ operations the Dirac spinor $\Psi $ transforms
as follows: 
\begin{eqnarray}
P\Psi \left( t,\vec{x}\right) P^{-1 } &=&\gamma ^{0}\Psi \left( t,-\vec{ x}%
\right)  \notag \\
C\Psi \left( t,\vec{x}\right) C^{-1 } &=&-i\gamma ^{2}\gamma ^{0}\bar{ \Psi}%
^{T}\left( t,\vec{x}\right)  \label{1} \\
T\Psi \left( t,\vec{x}\right) T^{-1 } &=&\gamma ^{1}\gamma ^{3}\Psi \left(
-t,\vec{x}\right)  \notag
\end{eqnarray}

The effect of transformations $C$, $P$ and $CP$ on various quantities that
appear in a gauge theory Lagrangian are given below:

\begin{equation*}
\begin{array}{ccccc}
\text{Transformation} \quad & \text{Scalar} \quad & \text{Pseudoscalar} \quad
& \text{ Vector} \quad & \text{Axial vector} \\ 
& \bar{\Psi}_{i}\Psi _{j} & i\bar{\Psi}_{i}\gamma _{5}\Psi _{j} & \bar{\Psi}
_{i}\gamma ^{\mu }\Psi _{j} & \bar{\Psi}_{i}\gamma ^{\mu }\gamma ^{5}\Psi
_{j} \\ 
P & \bar{\Psi}_{i}\Psi _{j} & -i\bar{\Psi}_{i}\gamma _{5}\Psi _{j} & \eta
\left( \mu \right) \bar{\Psi}_{i}\gamma ^{\mu }\Psi _{j} & -\eta \left( \mu
\right) \bar{\Psi}_{i}\gamma ^{\mu }\gamma ^{5}\Psi _{j} \\ 
C & \bar{\Psi}_{j}\Psi _{i} & i\bar{\Psi}_{j}\gamma _{5}\Psi _{i} & -\bar{
\Psi}_{j}\gamma ^{\mu }\Psi _{i} & \bar{\Psi}_{j}\gamma ^{\mu }\gamma
^{5}\Psi _{i} \\ 
CP & \bar{\Psi}_{j}\Psi _{i} & -i\bar{\Psi}_{j}\gamma _{5}\Psi _{i} & -\eta
\left( \mu \right) \bar{\Psi}_{j}\gamma ^{\mu }\Psi _{i} & -\eta \left( \mu
\right) \bar{\Psi}_{j}\gamma ^{\mu }\gamma ^{5}\Psi _{i}%
\end{array}%
\end{equation*}

The vector bosons associated with the electroweak unification group $%
SU_{L}\left(2\right) \times U\left(1\right) $ transform under $CP$ as: 
\begin{eqnarray}
&&W_{\mu }^{\pm }\left( \vec{x},t\right) \overset{CP}{\rightarrow }-\eta
\left( \mu \right) W_{\mu }^{\mp }\left( -\vec{x},t\right)  \notag \\
&&Z_{\mu }\left( \vec{x},t\right) \overset{CP}{\rightarrow }-\eta \left( \mu
\right) Z_{\mu }\left( -\vec{x},t\right)  \label{2} \\
&&A_{\mu }\left( \vec{x},t\right) \overset{CP}{\rightarrow }-\eta \left( \mu
\right) A_{\mu }\left( -\vec{x},t\right)  \notag
\end{eqnarray}
where, 
\begin{equation*}
\eta \left( \mu \right)= 
\begin{cases}
+1, & \text{if $\mu$=0} \\ 
-1, & \text{if $\mu$=1,2,3}%
\end{cases}%
\end{equation*}

The Lagrangian transforms as: 
\begin{eqnarray*}
\mathcal{L}_{W} &=&\bar{\psi}_{i}\gamma ^{\mu }(1-\gamma ^{5})\psi
_{j}W_{\mu }^{+}+h.c. \\
&\overset{CP}{\rightarrow } &-\eta (\mu )\bar{\psi}_{j}\gamma ^{\mu
}(1-\gamma ^{5})\psi _{i}(-\eta (\mu ))W_{\mu }^{-}+h.c.
\end{eqnarray*}

Thus, the weak interaction Lagrangian in the Standard Model violates $C$ and 
$P$ but is $CP$-invariant.

It is instructive to discuss the restrictions imposed by $CPT$ invariance. $%
CPT$ invariance implies, 
\begin{eqnarray}
_{\text{out}}\left\langle f\left| \mathcal{L}\right| X\right\rangle &=&_{%
\text{out}}\left\langle f\left| \left( CPT\right) ^{-1}\mathcal{L}CPT\right|
X\right\rangle  \notag \\
&=&\eta _{T}^{x\ast }\eta _{T}^{f}\,\,_{\text{in}}\left\langle \tilde{f}
\left| \left( CP\right) ^{\dagger }\mathcal{L}^{\dagger }\left( CP\right)
^{-1\dagger }\right| X\right\rangle ^{\ast }  \notag \\
&=&\eta _{T}^{x\ast }\eta _{T}^{f}\left\langle X\left| \left( CP\right)
^{-1} \mathcal{L}\left( CP\right) \right| f\right\rangle _{\text{in}}  \notag
\\
&=&-\eta _{T}^{x\ast }\eta _{T}^{f}\eta _{CP}^{f}\left\langle \bar{X}\left| 
\mathcal{L}S_{f}\right| \bar{f}\right\rangle _{\text{out}}  \notag \\
&=&\eta _{f}\,\,_{\text{out}}\left\langle \bar{f}\left| S_{f}^{\dagger } 
\mathcal{L}^{\dagger }\right| \bar{X}\right\rangle ^{\ast }  \notag \\
&=&\eta _{f}\,\exp (2i\delta _{f})_{\text{out}}\left\langle \bar{f}\left| 
\mathcal{L}\right| \bar{X}\right\rangle ^{\ast }  \label{3}
\end{eqnarray}

Hence, we get: 
\begin{eqnarray}
_{\text{out}}\left\langle \bar{f}\left| \mathcal{L}\right| \bar{X}
\right\rangle &=&\eta _{f}\,\exp (2i\delta _{f})_{\text{out}}\left\langle
f\left| \mathcal{L}\right| X\right\rangle ^{\ast }  \notag \\
\bar{A}_{\bar{f}} &=&\eta _{f}\exp (2i\delta _{f})A_{f}^{\ast }  \label{4}
\end{eqnarray}

In deriving the above result, we have put $\tilde{f}=f$ where $\tilde{f}$
means that momenta and spin are reversed. Since we are in the rest frame of $%
X$, $T$ will reverse only magnetic quantum number and we can drop $\tilde{f}$%
. Further we have used,

\begin{equation}
CP\left| X\right\rangle =-\left| \bar{X}\right\rangle  \label{5}
\end{equation}

\begin{equation}
CP\left| f\right\rangle =\eta _{f}^{CP}\left| \bar{f}\right\rangle  \label{6}
\end{equation}

\begin{equation}
\left\vert f\right\rangle _{\text{in}}=S_{f}\left\vert f\right\rangle _{%
\text{out}}=\exp (2i\delta _{f})\left\vert f\right\rangle _{\text{in}}
\label{7}
\end{equation}

where $\delta _{f}$ is the strong interaction phase shift. If $CP$%
-invariance holds, then,

\begin{equation*}
_{\text{out}}\left\langle f\left| \mathcal{L}\right| X\right\rangle =_{\text{
out}}\left\langle \bar{f}\left| \mathcal{L}\right| \bar{X}\right\rangle 
\newline
\Rightarrow \bar{A}_{\bar{f}}=A_{f}.
\end{equation*}

Thus, the necessary condition for $CP$-violation is that the decay amplitude 
$A$ should be complex. In view of our discussion above, under $CP$ an
operator $O\left(\vec{x},t\right)$ is replaced by,

\begin{equation}
O\left( \vec{x},t\right) \rightarrow O^{\dagger }\left( -\vec{x},t\right)
\label{9}
\end{equation}

The effective Lagrangian has the structure ($\mathcal{L}^{\dagger}=\mathcal{L%
}$),

\begin{equation}
\mathcal{L}=aO+a^{\ast }O^{\dagger }  \label{10}
\end{equation}

Hence, $CP$-violation requires $a^{\ast}\neq a$. We now discuss the
implication of $CPT$ constraint with respect to $CP$ violation of weak
decays. The weak amplitude is complex; it contains the final state strong
phase $\delta _{f}$ and in addition it may also contain a weak phase $\phi$.
Taking out both these phases,

\begin{equation*}
A_{f}=\exp (i\phi )F_{f}=\exp (i\phi )\exp (i\delta _{f})\left| F_{f}\right|
\end{equation*}

$CPT$ (Eq. \eqref{4}) gives,

\begin{equation*}
\bar{A}_{\bar{f}}=\exp (2i\delta _{f})\exp (-i\phi )\exp (-i\delta
_{f})\left| F_{f}\right| =\exp (-i\phi )F_{f}
\end{equation*}

We conclude that the weak interaction Lagrangian in the Standard Model is \ $%
CP$ invariant and since $CP$ violation has been observed in hadronic sector
(only in $B, B_{s}$ and $K$ decays) and not in leptonic sector, it is a
consequence of mismatch between weak and mass eigenstates (i.e. the phases
in $CKM$ matrix) and/or the mismatch between $CP$-eigenstates,

\begin{equation}
\left| X_{1,2}^{0}\right\rangle =\frac{1}{\sqrt{2}}\left[ \left|
X^{0}\right\rangle \mp \left| \bar{X}^{0}\right\rangle \right] ;\text{ }
CP\left| X_{1,2}^{0}\right\rangle =\pm \left| X_{1,2}^{0}\right\rangle
\label{11}
\end{equation}

and the mass eigenstates i.e. $CP$-violation in the mass matrix. $CP$-
violation due to mass mixing and in the decay amplitude has been
experimentally observed in $K^{0}$ and $B_{d}^{0}$. For $B_{s}$ decays, the $%
CP$-violation in the mass matrix is not expected in the Standard Model. In
fact time dependent $CP$-violation asymmetry gives a clear way to observe
direct $CP$-violation in $B$ and $B_{s}$ decays.

If $CP$ is conserved, 
\begin{eqnarray*}
\left\langle X_{2}\left| H\right| X_{1}\right\rangle &=&\left\langle
X_{2}\left| \left( CP\right) ^{-1}H\left( CP\right) \right|
X_{1}\right\rangle \\
&=&-\left\langle X_{2}\left| H\right| X_{1}\right\rangle
\end{eqnarray*}

then,

\begin{equation*}
\left\langle X_{2}\left| H\right| X_{1}\right\rangle =0.
\end{equation*}

Thus $\left| X_{1}\right\rangle $ and $\left| X_{2}\right\rangle $ are also
mass eigenstates. They form a complete set (in units $\hbar=c=1$), 
\begin{eqnarray}
\left| \psi \left( t\right) \right\rangle &=&a\left( t\right) \left|
X_{1}\right\rangle +b\left( t\right) \left| X_{2}\right\rangle  \notag \\
i\frac{d\left| \psi \left( t\right) \right\rangle }{dt} &=&\left( 
\begin{array}{cc}
m_{1}-\frac{i}{2}\Gamma _{1} & 0 \\ 
0 & m_{2}-\frac{i}{2}\Gamma _{2}%
\end{array}
\right) \left| \psi \left( t\right) \right\rangle .  \label{2.1}
\end{eqnarray}

The solution is, 
\begin{eqnarray*}
a\left( t\right) &=&a\left( 0\right) \exp \left( -im_{1}t-\frac{1}{2}\Gamma
_{1}t\right) \\
b\left( t\right) &=&b\left( 0\right) \exp \left( -im_{2}t-\frac{1}{2}\Gamma
_{2}t\right)
\end{eqnarray*}

Suppose we start with the state $\left|X^{0}\right\rangle$, i.e.,

\begin{equation*}
\left| \psi \left( 0\right) \right\rangle =\left| X^{0}\right\rangle
\end{equation*}

Then we get, 
\begin{eqnarray}
\left| \psi \left( t\right) \right\rangle &=&\frac{1}{\sqrt{2}}\left[ \exp
\left( -im_{1}t-\frac{1}{2}\Gamma _{1}t\right) \left| X_{1}\right\rangle
\right.  \notag \\
&&+\left. \exp \left( -im_{2}t-\frac{1}{2}\Gamma _{2}t\right) \left|
X_{2}\right\rangle \right]  \notag \\
&=&\frac{1}{\sqrt{2}}\left\{ \left[ \exp \left( -im_{1}t-\frac{1}{2}\Gamma
_{1}t\right) \right. \right.  \notag \\
&&\left. +\exp \left( -im_{2}t-\frac{1}{2}\Gamma _{2}t\right) \right] \left|
X^{0}\right\rangle  \notag \\
&&-\left[ \exp \left( -im_{1}t-\frac{1}{2}\Gamma _{1}t\right) \right.  \notag
\\
&&\left. \left. -\exp \left( -im_{2}t-\frac{1}{2}\Gamma _{2}t\right) \right]
\left| \bar{X}^{0}\right\rangle \right\}  \label{2.2}
\end{eqnarray}

However, in $\left|X^{0}\right\rangle-\left|\bar{X}^{0}\right\rangle$ basis, 
\begin{eqnarray*}
|\psi (t)\rangle &=&a(t)|X^{0}\rangle +\bar{a}(t)|\bar{X}^{0}\rangle \\
\frac{i}{dt}|\psi (t)\rangle &=&M|\psi (t)\rangle
\end{eqnarray*}

the mass matrix $M$ is not diagonal and is given by, 
\begin{equation}
M=m-\frac{i}{2}\Gamma= \left( 
\begin{array}{cc}
m_{11}-\frac{i}{2}\Gamma _{11} & m_{12}-\frac{i}{2}\Gamma _{12} \\ 
m_{21}-\frac{i}{2}\Gamma _{21} & m_{22}-\frac{i}{2}\Gamma _{22}%
\end{array}
\right)  \label{2.3}
\end{equation}

Hermiticity of matrices $m_{\alpha \alpha ^{\prime }}$ and $\Gamma _{\alpha
\alpha ^{\prime }}$ gives ($\alpha=\alpha ^{\prime}=1,2$), 
\begin{eqnarray}
\left( m\right) _{\alpha \alpha ^{\prime }} &=&\left( m^{\dagger }\right)
_{\alpha \alpha ^{\prime }}=\left( m^{\ast }\right) _{\alpha ^{\prime
}\alpha }, \qquad \Gamma _{\alpha \alpha ^{\prime }}=\Gamma _{\alpha
^{\prime }\alpha }^{\ast }  \notag \\
m_{21} &=&m_{12\,}^{\ast } \qquad \Gamma _{21}=\Gamma _{12}^{\ast }
\label{2.4}
\end{eqnarray}

$CPT$ invariance gives,

\begin{equation}
\left\langle X^{0}\left| M\right| X^{0}\right\rangle =\left\langle \bar{X}
^{0}\left| M\right| \bar{X}^{0}\right\rangle  \notag
\end{equation}

\begin{equation*}
m_{11}=m_{22}, \qquad \Gamma _{11}=\Gamma _{22}
\end{equation*}

\begin{equation}
\left\langle \bar{X}^{0}\left| M\right| X^{0}\right\rangle=\left\langle \bar{%
X}^{0}\left| M\right| X^{0}\right\rangle \text{: identity}  \label{2.5}
\end{equation}

Diagonalization of mass matrix $M$ in eq. (\ref{2.3}) gives, 
\begin{eqnarray}
m_{11}-\frac{i}{2}\Gamma _{11}-pq &=&m_{1}-\frac{i}{2}\Gamma _{1}  \notag \\
m_{11}-\frac{i}{2}\Gamma _{11}+pq &=&m_{2}-\frac{i}{2}\Gamma _{2}
\label{2.6}
\end{eqnarray}

where, 
\begin{equation}
p^{2}=m_{12}-\frac{i}{2}\Gamma _{12}, \qquad q^{2}=m_{12}^{\ast }-\frac{i}{2}%
\Gamma _{12}^{\ast}  \label{2.7}
\end{equation}

The eigenstates are given by, 
\begin{equation*}
|X_{1,2}\rangle =\frac{1}{\sqrt{\left| p\right| ^{2}+\left| q\right| ^{2}}} %
\left[ p|X^{0}\rangle \mp q|\bar{X}^{0}\rangle \right]
\end{equation*}

\section{$K^{0}-\bar{K}^{0}$ Complex and $CP$--Violation in $K$-Decay}

Consider the process, 
\begin{equation*}
K^{0}\rightarrow \pi^{+}\pi^{-}\rightarrow \bar{K}^{0}, \qquad \left|\Delta
Y\right|=2
\end{equation*}

Thus, weak interaction can mix $K^{0}$ and $\bar{K}^{0}$, 
\begin{equation*}
\left\langle K^{0}\left| H\right| \bar{K}^{0}\right\rangle \neq 0.
\end{equation*}

Off diagonal matrix elements are not zero. Thus, $K^{0}$ and $\bar{K}^{0}$
cannot be mass eigenstates.

Select the phase: 
\begin{equation*}
CP\left| K^{0}\right\rangle =-\left| \bar{K}^{0}\right\rangle .
\end{equation*}

Define, 
\begin{eqnarray*}
\left| K_{1}^{0}\right\rangle &=&\frac{1}{\sqrt{2}}\left[ \left|
K^{0}\right\rangle -\left| \bar{K}^{0}\right\rangle \right] \\
\left| K_{2}^{0}\right\rangle &=&\frac{1}{\sqrt{2}}\left[ \left|
K^{0}\right\rangle +\left| \bar{K}^{0}\right\rangle \right]
\end{eqnarray*}
Choose: 
\begin{equation*}
CP\left| K_{1}^{0}\right\rangle =+\left| K_{1}^{0}\right\rangle \qquad
CP\left| K_{2}^{0}\right\rangle =-\left| K_{2}^{0}\right\rangle
\end{equation*}

where $K_{1}^{0}$ and $K_{2}^{0}$ are eigenstates of $CP$ with eigenvalues $%
+1$ and $-1$.

Assuming $CP$ conservation, 
\begin{equation}
\left\langle \bar{K}^{0}\left| M\right| K^{0}\right\rangle =\left\langle
K^{0}\left| M\right| \bar{K}^{0}\right\rangle  \label{2.8}
\end{equation}
\begin{equation*}
m_{21}=m_{12} \qquad \Gamma _{21}=\Gamma _{12}
\end{equation*}

where $m_{12}$ and $\Gamma _{12}$ are real. Thus, 
\begin{eqnarray}
pq &=&m_{12}-\frac{i}{2}\Gamma _{12}  \notag \\
m_{1} &=&m_{11}-m_{12},\qquad \Gamma _{1}=\Gamma _{11}-\Gamma _{12}  \notag
\\
m_{2} &=&m_{11}+m_{12},\qquad \Gamma _{2}=\Gamma _{11}+\Gamma _{12}  \notag
\\
\Delta m &=&m_{2}-m_{1}=2m_{12}, \\
\,\Delta \Gamma &=&\Gamma _{2}-\Gamma _{1}=2\Gamma _{12}  \label{2.10}
\end{eqnarray}

Since, 
\begin{equation*}
CP\left( \pi ^{+}\,\pi ^{-}\right) =\left( -1\right) ^{l}\left( -1\right)
^{l}=1
\end{equation*}

therefore, it is clear that, 
\begin{equation*}
K_{1}^{0}\longrightarrow \pi ^{+}\,\pi ^{-}
\end{equation*}

is allowed by $CP$ conservation.

However, experimentally it was found that long lived $K_{2}^{0}$ also decay
to $\pi ^{+}\,\pi ^{-}$ but with very small probability. Small $CP$ non
conservation can be taken into account by defining, 
\begin{eqnarray}
\left\vert K_{S}\right\rangle &=&\left\vert K_{1}^{0}\right\rangle
+\varepsilon \left\vert K_{2}^{0}\right\rangle  \notag \\
\left\vert K_{L}\right\rangle &=&\left\vert K_{2}^{0}\right\rangle
+\varepsilon \left\vert K_{1}^{0}\right\rangle  \label{2.11}
\end{eqnarray}

where $\varepsilon $ is a small number. Thus $CP$ non conservation manifests
itself by the ratio: 
\begin{eqnarray}
\eta _{+-} &=&\frac{A\left( K_{L}\rightarrow \pi ^{+}\,\pi ^{-}\right) }{
A\left( K_{S}\rightarrow \pi ^{+}\,\pi ^{-}\right) }=\varepsilon
\label{2.12} \\
\left| \eta _{+-}\right| &\simeq &\left( 2.286\pm 0.017\right) \times 10^{-3}
\notag
\end{eqnarray}

Now $CP$ non conservation implies, 
\begin{equation}
m_{12}\neq m_{12}^{\ast }, \qquad \Gamma _{12}\neq \Gamma _{12}^{\ast }.
\label{2.13}
\end{equation}

Since $CP$ violation is a small effect, therefore, 
\begin{equation}
\text{Im}m_{12}\ll \text{Re}m_{12} \qquad \text{Im}\Gamma _{12} \ll \text{Re}%
\Gamma _{12}.  \label{2.14}
\end{equation}

Further, if $CP$- violation arises from mass matrix, then, 
\begin{equation}
\Gamma _{12}=\Gamma _{12}^{\ast }.  \label{2.15}
\end{equation}

Thus, $CP$--violation can result by a small term $i\text{Im}m_{12}$ in the
mass matrix given in Eq. (\ref{2.1}), 
\begin{equation}
M=\left( 
\begin{array}{cc}
m_1-\frac i2\Gamma _1 & i\text{Im}m_{12} \\ 
-i\text{Im}m_{12} & m_2-\frac i2\Gamma _2%
\end{array}
\right) .  \label{2.16}
\end{equation}

Diagonalization gives, 
\begin{equation}
\varepsilon =\frac{i\text{Im}m_{12}}{\left( m_2-m_1\right) -i\left( \Gamma
_2-\Gamma _1\right) /2}.  \label{2.17}
\end{equation}

Then from Eq. (\ref{2.10}) up to first order, we get, 
\begin{eqnarray}
\Delta m &=&m_2-m_1\rightarrow m_{K_L}-m_{K_S}  \notag \\
&=&2\text{Re}m_{12}  \notag \\
\Delta \Gamma &=&\Gamma _2-\Gamma _1=\Gamma _L-\Gamma _S=2\Gamma _{12}
\label{2.18}
\end{eqnarray}

Eq. (\ref{2.2}) is unchanged, replace, 
\begin{equation*}
m_1 \rightarrow m_S, \qquad m_2\rightarrow m_L
\end{equation*}
\begin{equation*}
\Gamma _1 \rightarrow \Gamma _S, \qquad \Gamma _2\rightarrow \Gamma _L
\end{equation*}

Now, 
\begin{eqnarray}
\Delta m &=&m_L-m_S  \notag \\
\Delta \Gamma &=&\Gamma _L-\Gamma _S  \notag \\
\Gamma _S &=&\frac \hbar {\tau _S}=7.367\times 10^{-12}\text{ MeV},\,\,\, 
\notag \\
&&\left. \tau _S=\left( 0.8935\pm 0.0008\right) \times 10^{-10}\text{ s}
\right.  \notag \\
\Gamma _L &=&\frac \hbar {\tau _L}=1.273\times 10^{-14}\text{ MeV},\,\, 
\notag \\
&&\left. \,\tau _L=\left( 5.17\pm 0.04\right) \times 10^{-8}\text{ s}\right.
\notag \\
\Delta \Gamma &\simeq &-\Gamma _S  \notag \\
m_L &=&m+\frac 12\Delta m  \notag \\
m_S &=&m-\frac 12\Delta m  \label{2.19}
\end{eqnarray}

Hence from Eq. (\ref{2.2}),

\begin{equation}
\left| \psi \left( t\right) \right\rangle \approx e^{\frac{-i}2mt}\left\{ 
\begin{array}{c}
\left[ e^{\frac{-1}2\Gamma _St}e^{\frac i2\Delta mt}+e^{-\frac i2\Delta mt} %
\right] \left| K^0\right\rangle \\ 
-\left[ e^{\frac{-1}2\Gamma _St}e^{\frac i2\Delta mt}-e^{-\frac i2\Delta mt} %
\right] \left| \bar{K}^0\right\rangle%
\end{array}
\right\}  \label{2.20}
\end{equation}

Therefore, probability of finding $\bar{K}^0$ at time $t$ (recall that we
started with $K^0$),

\begin{eqnarray}
P\left( K^0\rightarrow \bar{K}^0,t\right) &=&\left| \left\langle \bar{K}
^0\left| {}\right. \psi \left( t\right) \right\rangle \right| ^2  \notag \\
&=&\frac 14\left( 1+e^{-\Gamma _St}-2e^{-\frac 12\Gamma _St}\cos \left(
\Delta m\right) t\right)  \notag \\
&=&\frac 14\left( 1+e^{-t/\tau _S}-2e^{-\frac 12t/\tau _S}\cos \left( \Delta
m\right) t\right)  \label{2.21}
\end{eqnarray}

If kaons were stable $(\tau _S\rightarrow \infty )$, then,

\begin{equation}
P\left( K^0\rightarrow \bar{K}^0,t\right) =\frac 12\left[ 1-\cos \left(
\Delta m\right) t\right]  \label{2.22}
\end{equation}

which shows that a state produced as pure $Y=1$ state at $t=0$ continuously
oscillates between $Y=1$ and $Y=-1$ state with frequency $\omega =\frac{
\Delta m}\hbar $ and period of oscillation,

\begin{equation}
\tau =\frac{2\pi }{\left( \Delta m/\hbar \right) }.  \label{2.23}
\end{equation}

Kaons, however, decay and their oscillations are damped.

By measuring the period of oscillation, $\Delta m$ can be determined.

\begin{equation}
\Delta m=m_{L}-m_{S}=\left( 3.489\pm 0.008\right) \times 10^{-12}\text{ MeV.}
\label{2.24}
\end{equation}

Such a small number is measured as a consequence of superposition principle
in quantum mechanics, 
\begin{eqnarray*}
\pi ^{-}p &\rightarrow &K^{0}\Lambda ^{0} \\
&&\left. ^{|}\!\!\!\longrightarrow \bar{K}^{0}p\rightarrow \pi ^{+}\Lambda
^{0}\right.
\end{eqnarray*}

$\pi ^{+}$ can only be produced by $\bar{K}^{0}$ in the final state. This
would give a clear indication of oscillation.

Coming back to $CP$-violation, 
\begin{eqnarray}
\varepsilon &=&\frac{i\text{Im}m_{12}}{\Delta m-i\Delta \Gamma /2}\qquad
\varepsilon =\left| \varepsilon \right| e^{i\phi _{\varepsilon }}
\label{2.25} \\
\tan \phi _{\varepsilon } &=&-2\Delta m/\Delta \Gamma =\Delta m/\Gamma
_{S}-\Gamma _{L}  \notag \\
&\approx &\frac{2\times 0.474\Gamma _{S}}{0.998\Gamma _{S}}  \notag \\
&\Rightarrow &\phi _{\varepsilon }=43.59\pm 0.05^{0}  \label{2.26} \\
\left| \epsilon \right| &=&(2.229\pm 0.012)\times 10^{-3}  \label{2.226}
\end{eqnarray}

So far we have considered $CP$-violation due to mixing in the mass matrix.
It is important to detect the $CP$-violation in the decay amplitude if any.
This is done by looking for a difference between $CP$-violation for the
final $\pi ^{0}\pi ^{0}$ state and that for $\pi ^{+}\pi ^{-}$. Now due to
Bose statistics, the two pions can be either in $I=0$ or $I=2$ states. Using
Clebsch-Gordon (CG) coefficients, 
\begin{eqnarray}
A\left( K^{0}\rightarrow \pi ^{+}\pi ^{-}\right) &=&\frac{1}{\sqrt{3}}\left[ 
\sqrt{2}A_{0}e^{i\delta _{0}}+A_{2}e^{i\delta _{2}}\right]  \notag \\
A\left( K^{0}\rightarrow \pi ^{0}\pi ^{0}\right) &=&\frac{1}{\sqrt{3}}\left[
A_{0}e^{i\delta _{0}}-\sqrt{2}A_{2}e^{i\delta _{2}}\right]  \label{2.27}
\end{eqnarray}

Now $CPT$-invariance gives, 
\begin{eqnarray}
A\left( \bar{K}^{0}\rightarrow \pi ^{+}\pi ^{-}\right) &=&\frac{1}{\sqrt{3}} %
\left[ \sqrt{2}A_{0}^{\ast }e^{i\delta _{0}}+A_{2}^{\ast }e^{i\delta _{2}} %
\right]  \notag \\
A\left( \bar{K}^{0}\rightarrow \pi ^{0}\pi ^{0}\right) &=&\frac{1}{\sqrt{3}} %
\left[ A_{0}^{\ast }e^{i\delta _{0}}-\sqrt{2}A_{2}^{\ast }e^{i\delta _{2}} %
\right]  \label{2.28}
\end{eqnarray}

The dominant decay amplitude is $A_{0}$ due to $\Delta I=1/2$ rule, $\left|
A_{2}/A_{0}\right| \simeq 1/22$. Using the Wu--Yang phase convention, we can
take $A_{0}$ to be real. Neglecting terms of order $\varepsilon \text{Re} 
\frac{ A_{2}}{A_{0}}$ and $\varepsilon \text{Im} \frac{A_{2}}{A_{0}}$, we
get, 
\begin{eqnarray}
\eta _{+-} &\equiv &\left| \eta _{+-}\right| e^{i\phi _{+-}}\simeq
\varepsilon +\varepsilon ^{\prime }  \notag \\
\eta _{00} &\equiv &\left| \eta _{00}\right| e^{i\phi _{00}}\simeq
\varepsilon -2\varepsilon ^{\prime }  \label{2.29}
\end{eqnarray}

where, 
\begin{equation}
\varepsilon ^{\prime }=\frac{i}{\sqrt{2}}e^{i\left( \delta _{2}-\delta
_{0}\right) } \text{Im}\frac{A_{2}}{A_{0}}  \label{2.30}
\end{equation}

Clearly $\varepsilon ^{\prime }$ measures the $CP$-violation in the decay
amplitude, since $CP$-invariance implies $A_{2}$ to be real.

After $35$ years of experiments at Fermilab and CERN, results have converged
on a definitive non-zero result for $\varepsilon ^{\prime }$, 
\begin{eqnarray}
R &=&\left| \frac{\eta _{00}}{\eta _{+-}}\right| ^{2}=\left| \frac{%
\varepsilon -2\varepsilon ^{\prime }}{\varepsilon +\varepsilon ^{\prime }}%
\right| ^{2},\qquad \varepsilon ^{\prime }\ll \varepsilon  \notag \\
&\simeq &\left| 1-\frac{3\varepsilon ^{\prime }}{\varepsilon }\right|
^{2}\simeq 1-6\text{Re}\left( \varepsilon ^{\prime }/\varepsilon \right) 
\notag \\
\text{Re}\left( \varepsilon ^{\prime }/\varepsilon \right) &=&\frac{1-R}{6}
\label{2.31} \\
&=&\left( 1.65\pm 0.26\right) \times 10^{-3}.  \label{2.32}
\end{eqnarray}

This is an evidence that although $\varepsilon ^{\prime }$ is a very small,
but $CP$-violation does occur in the decay amplitude. Further we note from
Eq. (\ref{2.30}), 
\begin{equation*}
\phi _{\varepsilon ^{\prime }}=\delta _{2}-\delta _{0}+\frac{\pi }{2}\approx
42.3\pm 1.5^{0}
\end{equation*}

where numerical value is based on an analysis of $\pi \pi $ scattering.

We now discuss the CP-asymmetry in leptonic decays of kaon. 
\begin{eqnarray*}
\frac{\Delta S}{\Delta Q} &=& 1 \\
K^{+} &\rightarrow &\pi ^{0}+l^{+}+\nu _{l} \\
K^{0} &\rightarrow &\pi ^{-}+l^{+}+\nu _{l}=f \\
\overline{K}^{0} &\rightarrow &\pi ^{+}+l^{-}+\overline{\nu }_{l}=f^{*}\text{
CPT} \\
\frac{\Delta S}{\Delta Q} &=&-1 \\
K^{0} &\rightarrow &\pi ^{+}+l^{-}+\overline{\nu }_{l}=g^{*} \\
\overline{K}^{0} &\rightarrow &\pi ^{-}+l^{+}+\nu _{l}=g\text{ CPT}
\end{eqnarray*}
\begin{eqnarray*}
A(K_{L}^{0} &\rightarrow &\pi ^{-}+l^{+}+\nu _{l})=\frac{1}{\sqrt{2}}%
[(1+\epsilon )f+(1-\epsilon )g] \\
A(K_{L}^{0} &\rightarrow &\pi ^{+}+l^{-}+\overline{\nu }_{l})=\frac{1}{\sqrt{%
2}}[(1+\epsilon )g^{*}+(1-\epsilon )f*]
\end{eqnarray*}

The CP-asymmetry parameter $\delta _{l}:$%
\begin{eqnarray*}
\delta _{l} &=&\frac{\Gamma (K_{L}^{0}\rightarrow \pi ^{-}l^{+}\nu
_{l})-\Gamma (K_{L}^{0}\rightarrow \pi ^{+}l^{-}\overline{\nu }_{l})}{\Gamma
(K_{L}^{0}\rightarrow \pi ^{-}l^{+}\nu _{l})+\Gamma (K_{L}^{0}\rightarrow
\pi ^{+}l^{-}\overline{\nu }_{l})} \\
&=&\frac{2\text{Re}\epsilon [\left| f\right| ^{2}-\left| g\right| ^{2}]}{%
\left| f\right| ^{2}+\left| g\right| ^{2}+(fg^{*}+f^{*}g)+O(\epsilon ^{2})}
\end{eqnarray*}

In the standard model $\frac{\Delta S}{\Delta Q}=-1$ transitions are not
allowed, thus $g=0$. Hence 
\begin{equation*}
\delta _{l} \approx 2\text{Re}\epsilon =(3.32\pm 0.06)10^{-3}[\text{Expt.
value}]
\end{equation*}

From Eq. (\ref{2.226}), we get 
\begin{equation*}
2\text{Re}\epsilon =2\left| \epsilon \right| \cos \phi _{\epsilon }
\end{equation*}
which gives on using expermintal values for $\left| \epsilon \right| $ and $%
\phi _{\epsilon }$%
\begin{equation*}
2\text{Re}\epsilon =(3.23\pm 0.02\times 10^{-3})
\end{equation*}
in agreement with the expermimental value for $\delta _{l}$

Finally we discuss CP-asymmetries for $K\rightarrow 3\pi $ decays. The
decays 
\begin{eqnarray*}
K^{+} &\rightarrow &\pi ^{+}\pi ^{0}\pi ^{0}\text{, }\pi ^{+}\pi ^{+}\pi ^{-}
\\
K^{0} &\rightarrow &\pi ^{+}\pi ^{-}\pi ^{0}\text{, }\pi ^{0}\pi ^{0}\pi ^{0}
\end{eqnarray*}
are partiy conserving decays i.e. the parity of the final state is $-1$. Now
the C-partiy of $\pi ^{0}$ and ($\pi ^{+}\pi ^{-})_{l^{\prime }}$ are given
by 
\begin{equation*}
C(\pi ^{0})=1,\text{ }C(\pi ^{+}\pi ^{-})=(-1)^{l^{\prime }}
\end{equation*}
and G-parity of pion is $-1.$ Thus 
\begin{eqnarray*}
CP|\pi ^{0}\pi ^{0}\pi ^{0} &>&=-|\pi ^{0}\pi ^{0}\pi ^{0}> \\
CP|\pi ^{+}\pi ^{-}\pi ^{0} &>&=(-1)^{l^{\prime }+1}|\pi ^{+}\pi ^{-}\pi
^{0}>
\end{eqnarray*}
Hence CP-conservation implies 
\begin{eqnarray*}
K_{2}^{0} &\rightarrow &\text{ }\pi ^{0}\pi ^{0}\pi ^{0}\text{ allowed.} \\
K_{1}^{0} &\rightarrow &\text{ }\pi ^{0}\pi ^{0}\pi ^{0}\text{ is forbidden.}
\\
K_{1}^{0} &\rightarrow &\pi ^{+}\pi ^{-}\pi ^{0}\text{ allowed if }l_{1}%
\text{ is odd.} \\
K_{2}^{0} &\rightarrow &\pi ^{+}\pi ^{-}\pi ^{0}\text{ allowed if }l_{1}%
\text{ is even.}
\end{eqnarray*}

Now G-partiy of three pions $\pi ^{+}\pi ^{-}\pi ^{0}:$%
\begin{eqnarray*}
G &=&C(-1)^{I}=(-1)^{l^{\prime }+I}=-1 \\
\text{Hence }l^{\prime } &=&\text{even},\text{ }I(\text{odd});\text{ }I=1,3
\\
l^{\prime } &=&\text{odd},\text{ }I(\text{even});\text{ }I=0,2
\end{eqnarray*}
Only $l^{\prime }=0$ decays are favored as the decays for $l^{\prime }>0$
are highly suppressed due to centrifugal barrier. Hence $K_{1}^{0}%
\rightarrow \pi ^{+}\pi ^{-}\pi ^{0}$ is highly suppressed. Thus we have to
take into account $I=1,3$ amplitudes viz $a_{1}$ and $a_{3}$. $I=3$
contribution is expected to be suppressed as it requires $\Delta I=\frac{5}{2%
}$ transition.

Hence CP-asymmetries of $K^{0}\rightarrow 3\pi $ decays are given by 
\begin{eqnarray*}
\eta _{000} &=&\frac{A(K_{s}\rightarrow \text{ }\pi ^{0}\pi ^{0}\pi ^{0})}{%
A(K_{L}\rightarrow \text{ }\pi ^{0}\pi ^{0}\pi ^{0})}=\frac{[i\text{Im}%
a_{1}+\epsilon \text{Re}a_{1}]}{\text{Re}a_{1}+i\epsilon \text{Im}a_{1}} \\
&\approx &\epsilon +i\frac{\text{Im} a_{1}}{\text{Re} a_{1}} \\
\eta _{+-0} &=&\frac{A(K_{s}\rightarrow \pi ^{+}\pi ^{-}\pi ^{0})}{%
A(K_{L}\rightarrow \pi ^{+}\pi ^{-}\pi ^{0})}\approx \epsilon +i\frac{\text{%
Im} a_{1}}{\text{Re}a_{1}}=\eta _{000}
\end{eqnarray*}

\section{$B^{0}-\bar{B}^{0}$ Complex}

For $B_{q}^{0}$ (q=d or s) we show below that both $m_{12}$ and $\Gamma
_{12} $ have the same phase. Thus, 
\begin{eqnarray}
m_{12} &=&\left| m_{12}\right| e^{-2i\phi _{M}}  \notag \\
\Gamma _{12} &=&\left| \Gamma _{12}\right| e^{-2i\phi _{M}}  \label{3.1} \\
\left| \Gamma _{12}\right| &\ll &\left| m_{12}\right|  \notag \\
p^{2} &=&e^{-2i\phi _{M}}\left[ \left| m_{12}\right| -i\left| \Gamma
_{12}\right| \right] \simeq \left| m_{12}\right| e^{-2i\phi _{M}}  \notag \\
q^{2} &=&e^{+2i\phi _{M}}\left[ \left| m_{12}\right| -i\left| \Gamma
_{12}\right| \right] \simeq \left| m_{12}\right| e^{2i\phi _{M}}  \label{3.2}
\\
2pq &=&2\left| m_{12}\right| =(m_{2}-m_{1})-\frac{i}{2}\left( \Gamma
_{2}-\Gamma _{1}\right)  \notag \\
\Rightarrow \Delta m_{B}&=&\frac{\left( m_{2}-m_{1}\right) }{2}=\left|
m_{12}\right|  \label{3.3} \\
\Delta \Gamma &=&\Gamma _{2}-\Gamma _{1}=0  \notag
\end{eqnarray}

The above equations follow from the fact that, 
\begin{equation*}
m_{12}-i\Gamma _{12}=\langle \bar{B}_{q}^{0}\left| H_{eff}^{\Delta
B=2}\right| B_{q}^{0}\rangle
\end{equation*}

$H_{eff}^{\Delta B=2}$ induces particle-antiparticle transition. For $\Delta
m_{12},$ $H_{eff}^{\Delta B=2}$ arises from the box diagram as shown in Fig.
2, where the dominant contribution comes out from the $t-$quark. Thus, 
\begin{equation*}
m_{12}\varpropto (V_{tb})^{2}(V_{tq}^{\ast })^{2}m_{t}^{2}
\end{equation*}


Now, 
\begin{equation*}
\Gamma _{12}\varpropto \sum_{f}\langle \bar{B}^{0}\left| H_{W}\right|
f\rangle \langle f\left| H_{W}\right| B^{0}\rangle
\end{equation*}

where the sum is over all the final states which contribute to both $%
B_{q}^{0}$ and $\bar{B}_{q}^{0}$ decays. Thus, 
\begin{equation*}
\Gamma _{12}\varpropto \left( V_{cb}V_{cq}^{\ast }+V_{ub}V_{uq}^{\ast
}\right) ^{2}m_{b}^{2}\propto (V_{tb})^{2}(V_{tq}^{\ast })^{2}m_{b}^{2}
\end{equation*}

Hence we have the result that, 
\begin{equation*}
\frac{\left| \Gamma _{12}\right| }{\left| m_{12}\right| }\sim \frac{%
m_{b}^{2} }{m_{t}^{2}}
\end{equation*}

Now $B_{d}^{0}\rightarrow \bar{B}_{d}^{0}$ transition: 
\begin{equation*}
\left( V_{tb}\right) ^{2}\left( V_{td}^{\ast }\right) ^{2}=A^{2}\lambda ^{6} 
\left[ \left( 1+\rho \right) ^{2}+\eta ^{2}\right] e^{2i\beta }
\end{equation*}

Hence, 
\begin{equation*}
m_{12}=\left| m_{12}\right| e^{2i\beta}, \qquad \Gamma_{12}=\left|
\Gamma_{12}\right| e^{2i\beta}, \qquad \phi _{M}=-\beta
\end{equation*}

On the other hand, $B_{s}^{0}\rightarrow \bar{B}_{s}^{0}$ transition:

\begin{equation}
\left( V_{tb}\right) ^{2}\left( V_{ts}^{\ast }\right) ^{2}=\left|
V_{ts}\right| ^{2}\approx A^{2}\lambda ^{4}  \label{3.26}
\end{equation}

\begin{equation}
m_{12}=\left| m_{12}\right|, \qquad \Gamma _{12}=\left| \Gamma _{12}\right|
\label{3.27}
\end{equation}

\begin{equation}
\phi _{M}=0  \label{3.28}
\end{equation}

Also we have,

\begin{equation*}
\frac{\Delta m_{B_{s}}}{\Delta m_{B_{d}}}=\frac{\left| m_{12}\right| _{s}}{%
\left| m_{12}\right| _{d}}=\frac{1}{\lambda ^{2}\left[ \left( 1+\rho \right)
^{2}+\eta ^{2}\right] }\xi
\end{equation*}

where $\xi $ is $SU(3)$ breaking parameter.

Hence the mass eigenstates $B_{L}^{0}$ and $B_{H}^{0}$ can be written as: 
\begin{eqnarray}
\left| B_{L}^{0}\right\rangle &=&\frac{1}{\sqrt{2}}\left[ \left|
B^{0}\right\rangle -e^{2i\phi _{M}}\left| \bar{B}^{0}\right\rangle \right]
\quad CP=+1, \phi_{M}\rightarrow 0  \label{3.4} \\
\left| B_{H}^{0}\right\rangle &=&\frac{1}{\sqrt{2}}\left[ \left| \bar{B}
^{0}\right\rangle +e^{2i\phi _{M}}\left| B^{0}\right\rangle \right] \quad
CP=-1, \phi_{M}\rightarrow 0  \label{3.5}
\end{eqnarray}

In this case, $CP$ violation occurs due to phase factor $e^{2i\phi _{M}}$ in
the mass matrix.

Now one gets (from Eq. (\ref{2.2})), using Eqs.(\ref{3.3}), (\ref{3.4}) and (%
\ref{3.5}), 
\begin{eqnarray}
\left| B^{0}\left( t\right) \right\rangle &=&e^{-imt}e^{-\frac{1}{2}\Gamma
t}\left\{ \cos \left( \frac{\Delta m}{2}t\right) \left| B^{0}\right\rangle
\right.  \notag \\
&&\left. -ie^{+2i\phi _{M}}\sin \left( \frac{\Delta m}{2}t\right) \left| 
\bar{B}^{0}\right\rangle \right\}  \label{3.8}
\end{eqnarray}

Similarly we get, 
\begin{eqnarray}
\left| \bar{B}^{0}\left( t\right) \right\rangle &=&-e^{-imt}e^{-\frac{1}{2}
\Gamma t}\left\{ \cos \left( \frac{\Delta m}{2}t\right) \left| \bar{B}
^{0}\right\rangle \right.  \notag \\
&&\left. -ie^{-2i\phi _{M}}\sin \left( \frac{\Delta m}{2}t\right) \left|
B^{0}\right\rangle \right\}  \label{3.9}
\end{eqnarray}

Suppose we start with $B^{0}$ viz $|B^{0}\left( 0\right)
\rangle=|B^{0}\rangle ,$ the probabilities of finding $\bar{B}^{0}$ and $%
B^{0}$ at time $t$ is given by, 
\begin{eqnarray*}
P\left( B^{0}\rightarrow \bar{B}^{0},t\right) &=&\left| \langle \bar{B}
^{0}|B^{0}\left( t\right) \rangle \right| ^{2} \\
&=&\frac{1}{2}e^{-\Gamma t}\left( 1-\cos (\Delta m\right) t) \\
P\left( B^{0}\rightarrow B^{0},t\right) &=&\left| \langle B^{0}|B^{0}\left(
t\right) \rangle \right| ^{2} \\
&=&\frac{1}{2}e^{-\Gamma t}\left( 1+\cos (\Delta m\right) t)
\end{eqnarray*}

These are equations of a damped harmonic oscillator, the angular frequency
of which is, 
\begin{equation*}
\omega =\frac{\Delta m}{\hslash }
\end{equation*}

We define the mixing parameter, 
\begin{eqnarray*}
r &=&\frac{\int_{0}^{T}\left| \langle \bar{B}^{0}|B^{0}\left( t\right)
\rangle \right| ^{2}dt}{\int_{0}^{T}\left| \langle B^{0}|B^{0}\left(
t\right) \rangle \right| ^{2}dt}=\frac{\chi }{1-\chi } \\
&\xrightarrow{T\rightarrow \infty} & \frac{\left( \Delta m/\Gamma \right)
^{2}}{2+\left( \Delta m/\Gamma \right) ^{2}}=\frac{x^{2}}{2+x^{2}}
\end{eqnarray*}

Experimentally, for $B_{d}^{0}$ and $B_{s}^{0}$, 
\begin{eqnarray*}
\Delta m_{B_{d}^{0}} &=&(0.507\pm 0.005)\times 10^{-12}\hbar
s^{-1}=(3.337\pm 0.033)\times 10^{-10}\text{MeV} \\
\Delta m_{B_{s}^{0}} &=&(17.77\pm 0.10\pm 0.007)\times 10^{-12}\hbar s^{-1}
=(1.17\pm 0.01)\times 10^{-10}\text{MeV} \\
x_{d} &=&\left( \frac{\Delta m_{B_{d}^{0}}}{\Gamma _{B_{d}^{0}}}\right)
=0.77\pm 0.008 \\
x_{s} &=&\left( \frac{\Delta m_{B_{s}^{0}}}{\Gamma _{B_{s}^{0}}}%
\right)=26.05\pm 0.25
\end{eqnarray*}

From Eq. (\ref{3.8}) and (\ref{3.9}), the decay amplitudes for, 
\begin{eqnarray}
B^{0}\left( t\right) &\rightarrow &f\,\,\,\,\,\,\,\,\,\,\,\,\,A_{f}\left(
t\right) =\left\langle f\left| H_{w}\right| B^{0}\left( t\right)
\right\rangle  \notag \\
\bar{B}^{0}\left( t\right) &\rightarrow &\bar{f}\,\,\,\,\,\,\,\,\,\,\,\, 
\bar{A}_{\bar{f}}\left( t\right) =\left\langle \bar{f}\left| H_{w}\right| 
\bar{B}^{0}\left( t\right) \right\rangle  \label{3.10}
\end{eqnarray}

are given by, 
\begin{eqnarray}
\,\,\,A_{f}\left( t\right) &=&e^{-imt}e^{-\frac{1}{2}\Gamma t}\left\{ \cos
\left( \frac{\Delta m}{2}t\right) A_{f}\right.  \notag \\
&&\,\,\,\,\,\,\,\,\left. -ie^{+2i\phi _{M}}\sin \left( \frac{\Delta m}{2}
t\right) \bar{A}_{\bar{f}}\right\}  \label{3.11} \\
\bar{A}_{\bar{f}}\left( t\right) &=&e^{-imt}e^{-\frac{1}{2}\Gamma t}\left\{
\cos \left( \frac{\Delta m}{2}t\right) \bar{A}_{\bar{f}}\right.  \notag \\
&&\left. -ie^{-2i\phi _{M}}\sin \left( \frac{\Delta m}{2}t\right) A_{\bar{f}
}\right\} .  \label{3.12}
\end{eqnarray}

From Eqs.(\ref{3.11}) and (\ref{3.12}), we get for the decay rates, 
\begin{eqnarray}
\Gamma _{f}(t) &=&e^{-\Gamma t}\left[ 
\begin{array}{c}
\frac{1}{2}\left( \left| A_{f}\right| ^{2}+\left| \bar{A}_{f}\right|
^{2}\right) +\frac{1}{2}\left( \left| A_{f}\right| ^{2}-\left| \bar{A}
_{f}\right| ^{2}\right) \cos \Delta mt \\ 
-\frac{i}{2}\left( 2i\text{Im}e^{2i\phi _{M}}A_{f}^{\ast } \bar{A}%
_{f}\right) \sin \Delta mt%
\end{array}
\right]  \label{I} \\
\bar{\Gamma}_{\bar{f}}(t) &=&e^{-\Gamma t}\left[ 
\begin{array}{c}
\frac{1}{2}\left( \left| A_{\bar{f}}\right| ^{2}+\left| \bar{A}_{\bar{f}
}\right| ^{2}\right) -\frac{1}{2}\left( \left| A_{\bar{f}}\right|
^{2}-\left| \bar{A}_{\bar{f}}\right| ^{2}\right) \cos \Delta mt \\ 
+\frac{i}{2}\left( 2i\text{Im}e^{2i\phi _{M}}A_{\bar{f}}^{\ast }\bar{A}_{%
\bar{f}}\right) \sin \Delta mt%
\end{array}
\right]  \label{II}
\end{eqnarray}

for $\Gamma _{\bar{f}}$ and $\bar{\Gamma}_{f}$ change $f\rightarrow \bar{f}$
and $\bar{f}\rightarrow f$ in $\Gamma _{f}$ and $\bar{\Gamma}_{\bar{f}}$
respectively.

As a simple application of the above equations, consider the semi-leptonic
decays of $B^{0}$, 
\begin{eqnarray*}
B^{0} &\rightarrow &l^{+}\nu X^{-}:f\text{ \ for example }X^{-}=D^{-} \\
\bar{B}^{0} &\rightarrow &l^{-}\bar{\nu}X^{+}:\bar{f}\text{ \ for example }%
X^{+}=D^{+}
\end{eqnarray*}

In the standard model, $\bar{B}^{0}$ decay into $l^{+}\nu X^{-}$ and $B^{0}$
decay into $l^{-}\bar{\nu}X^{+}$ is forbidden. Thus, 
\begin{eqnarray*}
\bar{A}_{f}&=&0, \qquad A_{\bar{f}}=0 \\
\Gamma _{f}(t)&=&e^{-\Gamma t}\frac{1}{2}\left| A_{f}\right|
^{2}\left(1+\cos \Delta mt\right) \\
\Gamma _{\bar{f}}(t)&=&e^{-\Gamma t}\frac{1}{2}\left| \bar{A}_{\bar{f}
}\right| ^{2}\left( 1-\cos \Delta mt\right), \qquad \because \left| \bar{A}_{%
\bar{f}}\right| =\left| A_{f}\right|
\end{eqnarray*}

Hence, 
\begin{equation*}
\delta =\frac{\int_{0}^{\infty }\Gamma _{\bar{f}}(t)dt}{\int_{0}^{\infty}%
\Gamma _{f}(t)dt}=\frac{x_{d}^{2}}{2+x_{d}^{2}}=r_{d}
\end{equation*}

Non zero value of $\delta $ would indicate mixing. If, however, $\bar{A}%
_{f}\neq 0$ and $A_{\bar{f}}\neq 0$ due to some exotic mechanism, then $%
\delta \neq 0$ even without mixing. Thus 
\begin{eqnarray*}
\frac{\Gamma \left( \mu ^{-}X^{+}\right) }{\Gamma \left( \mu
^{+}X^{-}\right) +\Gamma \left( \mu ^{-}X^{+}\right) } &=&\frac{r_{d}}{%
1+r_{d}}=\chi _{d} \\
&=&0.172\pm 0.010\text{ (Expt value)}
\end{eqnarray*}

which gives, 
\begin{equation*}
x_{d}=0.723\pm 0.032
\end{equation*}

\section{$CP$-Violation in $B$-Decays}

\textbf{Case-I:} 

\begin{equation*}
|\bar{f}\rangle =CP|f\rangle =|f\rangle
\end{equation*}

For this case we get, from Eqs. (\ref{3.11}) and (\ref{3.12}), 
\begin{eqnarray}
\mathcal{A}_{f}\left( t\right) &=&\frac{\Gamma _{f}\left( t\right) -\bar{
\Gamma}_{f}\left( t\right) }{\Gamma _{f}\left( t\right) +\bar{\Gamma}
_{f}\left( t\right) }=\cos \left( \Delta mt\right) \left( \left|
A_{f}\right| ^{2}-\left| \bar{A}_{f}\right| ^{2}\right)  \notag \\
&&-i\sin \left( \Delta mt\right) \left( e^{2i\phi _{M}}A_{f}^{\ast }\bar{A}
_{f}-e^{-2i\phi _{M}}A_{f}\bar{A}_{f}^{\ast }\right) /\left(\left|
A_{f}\right| ^{2}+\left| \bar{A}_{f}\right| ^{2} \right)  \label{3.13} \\
&=&\cos \left( \Delta mt\right) C_{\pi \pi }+\sin \left( \Delta mt\right)
S_{\pi \pi }  \label{3.14}
\end{eqnarray}

where, 
\begin{equation*}
C_{\pi \pi }=\frac{1-\left| \bar{A}_{f}\right| ^{2}/\left| A_{f}\right|^{2}}{%
1+\left| \bar{A}_{f}\right| ^{2}/\left| A_{f}\right| ^{2}}=\frac{1-\left|
\lambda \right| ^{2}}{1+\left| \lambda \right| ^{2}} \qquad \lambda=\frac{%
\bar{A}_{f}}{A_{f}}
\end{equation*}

This is the direct $CP$ violation and, 
\begin{equation*}
S_{\pi\pi}=\frac{2\text{Im}\left(e^{2i\phi_{M}}\lambda\right)}{1+\left|
\lambda \right| ^{2}}
\end{equation*}

is the mixing induced $CP$-violation.

If the decay proceeds through a single diagram (for example tree graph), $%
\bar{A}_{f}/A_{f}$ is given by (see Eqs. (15) and (16)), 
\begin{equation*}
\lambda =\frac{\bar{A}_{f}}{A_{f}}=\frac{e^{i\left( \phi +\delta _{f}\right)
}}{e^{i\left( -\phi +\delta _{f}\right) }}=e^{2i\phi }
\end{equation*}

where $\phi $ is the weak phase in the decay amplitude. Hence from Eq. (\ref%
{3.13}), we obtain, 
\begin{equation}
\mathcal{A}_{f}(t)=\sin \left( \Delta mt\right) \sin \left( 2\phi _{M}+2\phi
\right)
\end{equation}

In particular for the decay, 
\begin{equation*}
B^{0}\rightarrow J/\psi \,K_{s},\,\,\,\,\phi =0
\end{equation*}

we obtain, 
\begin{equation}
\mathcal{A}_{\psi K_{s}}\left( t\right) =\sin \left( 2\phi_{M} \right) \sin
\left( \Delta mt\right) =-\sin 2\beta \sin (\Delta mt)  \label{3.15}
\end{equation}

and, 
\begin{eqnarray}
\mathcal{A}_{\psi K_{s}} &=&\frac{\int_{0}^{\infty }\left[ \Gamma _{f}\left(
t\right) -\bar{\Gamma}_{f}\left( t\right) \right] dt}{\int_{0}^{\infty } %
\left[ \Gamma _{f}\left( t\right) +\bar{\Gamma}_{f}\left( t\right) \right]
dt }  \notag \\
\mathcal{A}_{\psi K_{s}} &=&-\sin \left( 2\beta \right) \,\,\frac{\left(
\Delta m/\Gamma \right) }{1+\left( \Delta m/\Gamma \right) ^{2}}
\label{3.16} \\
\text{Experiment} &:&\left( \frac{\Delta m}{\Gamma }\right)
_{B_{d}^{0}}=0.776\pm 0.008  \label{3.17}
\end{eqnarray}

$\mathcal{A}_{\psi K_{s}}$ has been experimentally measured. It gives, 
\begin{equation*}
\sin 2\beta = 0.678\pm 0.025
\end{equation*}

Corresponding to the decay $B^{0}\rightarrow J/\psi \,K_{s}$, we have the
decay $B_{s}^{0}\rightarrow J/\psi \,\phi .$ Thus for this decay 
\begin{eqnarray*}
\mathcal{A}_{J/\psi \phi }^{(t)} &=&-\sin 2\beta _{s}\sin (\Delta m_{B
_{s}}t) \\
\mathcal{A}_{J/\psi \phi } &=&-\sin 2\beta _{s}\frac{(\Delta m_{B
_{s}}/\Gamma _{s})}{1+(\Delta m_{B _{s}}/\Gamma _{s})^{2}}
\end{eqnarray*}

In the standard model, $\beta _{s}=0,$ $\mathcal{A}_{J/\psi \phi }=0.$

This is an example of $CP$-violation in the mass matrix. We now discuss the
direct $CP$-violation.

Direct $CP$-violation in $B$ decays involves the weak phase in the decay
amplitude. The reason for this being that necessary condition for direct $CP$
-violation is that decay amplitude should be complex as discussed in section
1. But this is not sufficient because in the limit of no final state
interactions, the direct $CP$-violation in $B\rightarrow f$, $\bar{B}
\rightarrow \bar{f}$ decay vanishes. To illustrate this point, we discuss
the decays $\bar{B}^{0}\rightarrow \pi ^{+}\pi ^{-}$. The main contribution
to this decay is from tree graph (see Fig. 3); But this decay can also
proceed via the penguin diagram (see Fig. 4).

The contribution of penguin diagram can be written as 
\begin{equation}
P=V_{ub}V_{ud}^{\ast }f\left( u\right) +V_{cb}V_{cd}^{\ast }f\left( c\right)
+V_{tb}V_{td}^{\ast }f\left( t\right)  \label{4.1}
\end{equation}

where $f\left( u\right) $, $f\left( c\right) $ and $f\left( d\right) $
denote the contributions of $u$, $c$ and $t$ quarks in the loop. Now using
the unitarity equation (\ref{07}), we can rewrite Eq. (\ref{4.1}) as, 
\begin{eqnarray}
P_{c} &=&V_{ub}V_{ud}^{\ast }\left( f\left( u\right) -f\left( t\right)
\right) +V_{cb}V_{cd}^{\ast }\left( f\left( c\right) -f\left( t\right)
\right)  \label{4.2} \\
\text{or }P_{t} &=&V_{ub}V_{ud}^{\ast }\left( f\left( u\right) -f\left(
c\right) \right) + V_{tb}V_{td}^{\ast }\left( f\left( t\right) -f\left(
c\right) \right)  \notag
\end{eqnarray}

Due to loop integration $P$ is suppressed relative to $T$ but still its
contribution is not negligible. The first part of Eq. (\ref{4.2}) has the
same CKM matrix elements as for the tree graph, so we can absorb it in the
tree graph. Hence we can write (with $f=\pi^{+}\pi^{-}$),

\begin{equation}
\bar{A}_{f}=A\left( \bar{B}^{0}\rightarrow \pi ^{+}\pi ^{-}\right) =\left|
T\right|e^{i\left( -\gamma +\delta _{T}\right) }+\left| P\right|e^{i\left(
\phi +\delta _{P}\right) }  \label{4.3}
\end{equation}

where $\delta _{T}$ and $\delta _{P}$ are strong interaction phases which
have been taken out so that $T$ and $P$ are real. $\phi $ is the weak phase
in Penguin graph. $CPT$ invariance gives,

\begin{equation}
A_{f}\equiv {}A\left( B^{0}\rightarrow \pi ^{+}\pi ^{-}\right) =\left|
T\right|e^{-i\left(- \gamma -\delta _{T}\right) }+\left| P\right|e^{-i\left(
\phi -\delta _{P}\right) }.  \label{4.4}
\end{equation}

Hence direct $CP$--violation asymmetry is given by, 
\begin{eqnarray}
A_{CP} &=&\frac{-\Gamma \left( B^{0}\rightarrow \pi ^{+}\pi ^{-}\right)
+\Gamma \left( \bar{B}^{0}\rightarrow \pi ^{+}\pi ^{-}\right) }{\Gamma
\left( B^{0}\rightarrow \pi ^{+}\pi ^{-}\right) +\Gamma \left( \bar{B}%
^{0}\rightarrow \pi ^{+}\pi ^{-}\right) }  \notag \\
&=&-\frac{1-\left\vert \lambda \right\vert ^{2}}{1+\left\vert \lambda
\right\vert ^{2}}  \notag \\
&=&\frac{-2r\sin \delta \sin \left( \phi +\gamma \right) }{1+2r\cos \delta
\cos \left( \gamma +\phi \right) +r^{2}}  \notag
\end{eqnarray}

where ($F_{\text{CKM}}$ = CKM factor), 
\begin{equation*}
\delta =\delta _{P}-\delta _{T} \qquad r \rightarrow F_{\text{CKM}}\frac{%
\left| P\right|}{\left| T\right|}
\end{equation*}

For the time dependent $CP$-asymmetry for $B^{0}\rightarrow \pi ^{+}\pi ^{-}$
decay we obtain from Eqs. (\ref{3.13}) and (\ref{4.3}), 
\begin{subequations}
\begin{equation}
\mathcal{A}(t)=C_{\pi \pi }(\cos \Delta mt)+S_{\pi \pi }(\sin \Delta mt),
\label{4.7a}
\end{equation}

where the direct CP--violation parameter $C_{\pi\pi}$ and the mixing induced
parameter $S_{\pi\pi}$ are given by, 
\begin{equation}
S_{\pi \pi }=\frac{2\text{Im}[e^{2i\phi _{M}}\lambda ]}{1+|\lambda |^{2}}=- 
\frac{\sin \left( 2\beta +2\gamma \right) +2r\cos \delta \sin \left( 2\beta
+\gamma -\phi \right) +r^{2}\sin (2\beta -2\phi )}{1+2r\cos \delta \cos
(\gamma +\phi )+r^{2}}  \label{4.7b}
\end{equation}

\begin{equation}
C_{\pi \pi }=-A_{CP}  \label{4.7c}
\end{equation}

As discussed above, we have two choices in selecting the Penguin
contribution.

For the first choice, 
\end{subequations}
\begin{equation*}
\phi =\pi ,\quad F_{\text{CKM}}=\frac{\left\vert V_{cb}\right\vert
\left\vert V_{cd}\right\vert }{\left\vert V_{ub}\right\vert \left\vert
V_{ud}\right\vert }=\frac{1}{\left( \bar{\rho}^{2}+\bar{\eta}^{2}\right)
^{1/2}}
\end{equation*}

\begin{equation*}
r=\frac{1}{R_{b}}\frac{\left\vert P_{C}\right\vert }{\left\vert T\right\vert 
}
\end{equation*}

For this case we have, 
\begin{eqnarray*}
C_{\pi \pi } &=&\frac{2r\sin \delta \sin \gamma }{1+2r\cos \delta \cos
\gamma +r^{2}} \\
S_{\pi \pi } &=&-\frac{\sin (2\beta +2\gamma )+2r\cos \delta \sin (2\beta
+\gamma )+r^{2}\sin 2\beta }{1+2r\cos \delta \cos \gamma +r^{2}} \\
&=&\frac{\sin (2\alpha )+2r\cos \delta \sin (\beta -\alpha )-r^{2}\sin
2\beta }{1+2r\cos \delta \cos \left( \alpha +\beta \right) +r^{2}}
\end{eqnarray*}

For the second choice, 
\begin{equation*}
\phi =\beta, \quad F_{\text{CKM}}=\frac{\left| V_{tb}\right| \left|
V_{td}\right| }{\left| V_{ub}\right| \left| V_{ud}\right| }\approx \frac{%
\sqrt{\left( 1-\bar{\rho} \right) ^{2}+\bar{\eta} ^{2}}}{\sqrt{\bar{\rho}%
^{2}+\bar{\eta^{2}}}}
\end{equation*}

\begin{equation*}
r=\frac{R_{t}}{R_{b}} \frac{\left| P_{t}\right|}{\left| T\right|}
\end{equation*}

So that in this case we get, 
\begin{eqnarray*}
C_{\pi \pi } &=&-A_{CP}=\frac{2r\sin \delta \sin \alpha }{1-2r\cos \delta
\cos \alpha +r^{2}} \\
S_{\pi \pi } &=& \frac{\sin 2\alpha-2r\cos \delta \sin \alpha}{1-2r\cos
\delta \cos \alpha +r^{2}}
\end{eqnarray*}

For $B^{+} \rightarrow \pi^{+} \pi^{-}$, $B^{0} \rightarrow \pi^{0} \pi^{0}$%
, the decay amplitudes are given by,

\begin{eqnarray*}
A_{00} &=&A(B^{0}\rightarrow \pi ^{0}\pi ^{0})=\frac{1}{\sqrt{2}}%
Te^{i\delta_{T}}e^{i\gamma}\left[-r_{c}e^{i\delta_{CT}} +
re^{-i(\phi+\gamma-\delta+\delta_{CT})}\right] \\
A_{+0} &=&A(B^{+}\rightarrow \pi ^{+}\pi ^{0})=\frac{1}{\sqrt{2}}%
Te^{i\delta_{T}}e^{i\gamma }\left[1+r_{C}e^{i\delta_{CT}}\right] \\
r_{C} &=&\frac{C}{T}, \qquad \delta_{CT}=\delta_{C} - \delta_{T}
\end{eqnarray*}

Hence for $B^{0}\rightarrow\pi^{0}\pi^{0}$, the $CP$--asymmetries are given
by 
\begin{eqnarray*}
C_{\pi ^{0}\pi ^{0}} &=&-A_{00}^{CP}=\frac{-2r/r_{C}\sin (\delta -
\delta_{CT})\sin (\gamma +\phi )}{1+r/r_{C}^{2}+2r/r_{C}\cos (\delta -
\delta _{CT})\cos (\gamma +\phi )} \\
S_{\pi ^{0}\pi ^{0}} &=&-\frac{\sin (2\beta +2\gamma)-2r/r_{C}\cos (\delta -
\delta _{CT})\sin (2\beta +\gamma -\phi) + r^{2}/r_{C}^{2}\sin (2\beta
-2\phi )}{1 + + r^{2}/r_{C}^{2} 2r/r_{C}\cos (\delta - \delta_{CT})\cos
(\gamma +\phi)}
\end{eqnarray*}

For the case $\phi=\beta$, we get, 
\begin{eqnarray*}
C_{\pi^{0}\pi^{0}} &=&-A_{00}^{CP}=\frac{-2r/r_{C}\sin (\delta -
\delta_{CT})\sin \alpha}{1+r^{2}/r_{C}^{2}-2r/r_{C}\cos (\delta - \delta
_{CT})\cos \alpha} \\
S_{\pi^{0}\pi^{0}} &=&-\frac{\sin 2\alpha - 2r/r_{C}\cos (\delta - \delta
_{CT})\sin \alpha}{1 + r^{2}/r_{C}^{2} - 2r/r_{C}\cos (\delta -
\delta_{CT})\cos \alpha}
\end{eqnarray*}

We end this section by considering the decays

\begin{equation*}
\bar{B}^{0}\rightarrow \phi K_{s}, \omega K_{s} \text{ and }\rho K_{s}
\end{equation*}

These decays satisfy the relations 
\begin{eqnarray*}
&&\left[ \frac{1}{\sqrt{2}}\left( 
\begin{array}{c}
\langle \rho ^{0}\bar{K}^{0}\left| H_{W}\left( s)\right) \right| \bar{B}
^{0}\rangle -\langle \omega \bar{K}^{0}\left| H_{W}\left( s)\right) \right| 
\bar{B}^{0}\rangle \\ 
-\langle \phi \bar{K}^{0}\left| H_{W}\left( s)\right) \right| \bar{B}
^{0}\rangle%
\end{array}
\right) \right] \\
&=&\left[ 
\begin{array}{c}
\frac{1}{2}\left( C-P+P_{EW}\right) -\frac{1}{2}\left( C+P+\frac{1}{3}
P_{EW}\right) \\ 
-(-P+\frac{1}{3}P_{EW})%
\end{array}
\right] =0
\end{eqnarray*}

where $C,P$ and $P_{EW}$ are color suppressed, penguin and electroweak
penguin amplitudes for these decay.

From the above equation, we obtain, 
\begin{eqnarray*}
\frac{\langle \Gamma \rangle _{\omega K}+\langle \Gamma \rangle _{\rho
^{0}K} }{\langle \Gamma \rangle _{\phi K}} &\approx &1 \\
\frac{S(\rho ^{0}K_{s})+S(\omega K_{s})}{2} &\approx &S(\phi K_{s})=-\sin
2\beta \\
C(\rho ^{0}K_{s}) &\approx &-C(\omega K_{s})
\end{eqnarray*}

where we have neglected the terms of the order $r^{2}$. The parameter $r$ is
defined below, 
\begin{eqnarray*}
\langle \rho ^{0}\bar{K}^{0}\left| H_{W}(s)\right| \bar{B}^{0}\rangle
&=&-\left| V_{cb}\right| \left| V_{cd}\right| \left| P\right| e^{i\delta
_{P}}\left[ 1-re^{i(\delta _{C}-\delta _{P})}e^{-i\gamma }\right] \\
r &=&\frac{\left| C\right| }{\left| P\right| }\lambda ^{2}R_{b} \\
\frac{\left| C\right| }{\left| P\right| } &\sim &\frac{a_{2}}{a_{4}}
\end{eqnarray*}

Assuming factorization for the electroweak penguin, we get from the above
equation an interesting sum rule, 
\begin{equation*}
f_{\rho }F_{1}^{B-K}(m_{\rho }^{2})-\frac{1}{3}f_{\omega
}F_{1}^{B-K}(m_{\omega }^{2})-\frac{2}{3}f_{\phi }F_{1}^{B-K}(m_{\phi}^{2})=0
\end{equation*}

Assuming $F_{1}(m_{\rho }^{2})=F_{1}(m_{\omega }^{2})=F_{1}(m_{\phi
}^{2})\approx F_{1}(1$ GeV$^{2})$, we get from the relation, 
\begin{equation*}
f_{\rho }-\frac{1}{3}f_{\omega }-\frac{2}{3}f_{\phi }=0
\end{equation*}

which is reminiscent of current algebra and spectral function sum rules of
1960's.

The above sum rule is very well satisfied by the experimental values, 
\begin{equation*}
f_{\rho}=\left( 209\pm 1\right) \text{MeV}, \qquad f_{\omega}=\left( 187\pm
3\right) \text{MeV}, \qquad f_{\phi}=\left( 221\pm 3\right) \text{ MeV.}
\end{equation*}

It is convenient to write, from Eqs.(\ref{I}) and (\ref{II}), the decay
rates in the following form, 
\begin{eqnarray}
&&\left[ \Gamma_{f}(t) -\bar{\Gamma}_{\bar{f}}(t) \right] +\left[ \Gamma_{%
\bar{f}}-\bar{\Gamma}_{f}(t) \right]  \notag \\
&=&e^{-\Gamma t}\left\{ \cos \Delta mt\left[ \left( \left| A_{f}\right|
^{2}-\left| \bar{A}_{\bar{f}}\right| ^{2}\right) +\left( \left| A_{\bar{f}
}\right| ^{2}-\left| \bar{A}_{f}\right| ^{2}\right) \right] \right.  \notag
\\
&&\left. +2\sin \Delta mt\left[ \text{Im}\left( e^{2i\phi _{M}}A_{f}^{\ast } 
\bar{A}_{f}\right) +\text{Im}\left( e^{2i\phi _{M}}A_{\bar{f}}^{\ast }\bar{A}
_{\bar{f}}\right) \right] \right\}  \notag \\
&&  \label{e1} \\
&&\left[\Gamma_{f}(t)+\bar{\Gamma}_{\bar{f}}(t) \right] - \left[\Gamma_{\bar{%
f}}(t)+\bar{\Gamma}_{f}(t) \right]  \notag \\
&=&e^{-\Gamma t}\left\{ \cos \Delta mt\left[ \left( \left| A_{f}\right|
^{2}+\left| \bar{A}_{\bar{f}}\right| ^{2}\right) -\left( \left| A_{\bar{f}
}\right| ^{2}+\left| \bar{A}_{f}\right| ^{2}\right) \right] \right.  \notag
\\
&&\left. +2\sin \Delta mt\left[ \text{Im}\left( e^{2i\phi _{M}}A_{f}^{\ast } 
\bar{A}_{f}\right) -\text{Im}\left( e^{2i\phi _{M}}A_{\bar{f}}^{\ast }\bar{A}
_{\bar{f}}\right) \right] \right\}  \notag \\
&&  \label{e2}
\end{eqnarray}

We now use the above equations to obtain some interesting results for the $%
CP $ asymmetries for B-decays.

\bigskip

\textbf{Case-II:} 

We first consider the case in which single weak amplitudes $A_{f}$ and $A_{%
\bar{f}}^{^{\prime}}$ with different weak phases describe the decays: 
\begin{eqnarray}
A_{f} &=&\langle f\left| \mathcal{L_{W}}\right| B^{0}\rangle =e^{i\phi}F_{f}
\notag \\
&&  \label{e3} \\
A_{\bar{f}}^{^{\prime}} &=&\langle \bar{f}\left| \mathcal{L_{W}^{^{\prime}}}%
\right| B^{0}\rangle =e^{i\phi^{^{\prime}}} F_{\bar{f}}^{^{\prime}}  \notag
\end{eqnarray}

$CPT$ gives, 
\begin{eqnarray}
\bar{A}_{\bar{f}} &=&\langle \bar{f}\left| \mathcal{L_{W}}\right| \bar{B}%
^{0}\rangle =e^{2i\delta _{f}}A_{f}^{\ast }  \notag \\
&&  \label{e4} \\
\bar{A}_{f}^{^{\prime}} &=&\langle f\left| \mathcal{L_{W}}%
^{^{\prime}}\right| \bar{B}^{0}\rangle =e^{2i\delta^{^{\prime}}_{\bar{f}}}A_{%
\bar{f}}^{\ast^{\prime}}  \notag
\end{eqnarray}

Note $\delta _{f}$ and $\delta _{\bar{f}}^{\prime }$ are strong phases; $%
\phi $ and $\phi ^{\prime }$ are weak phases. The states $|f>$ and $|%
\overline{f}>$ are C-conjugate of each other such as states $D^{(*)-}\pi
^{+}(D^{(*)+}\pi ^{-}),$ $D_{s}^{(*)-}K^{+}(D_{s}^{(*)+}K^{-}),$ $D^{-}\rho
^{+}(D^{+}\rho ^{-})$

Hence, we get from Eqs.(\ref{e1}), (\ref{e2}), (\ref{e3}) and (\ref{e4}), 
\begin{eqnarray}
\mathcal{A}\left( t\right) &\equiv &\frac{[\Gamma _{f}(t)-\bar{\Gamma}_{\bar{%
f}}(t)]+[\Gamma _{\bar{f}}(t)-\bar{\Gamma}_{f}(t)]}{[\Gamma _{f}(t)+\bar{%
\Gamma}_{\bar{f}}(t)]+[\Gamma _{\bar{f}}(t)+\bar{\Gamma}_{f}]}  \notag \\
&=&\frac{2\bigl|F_{f}\bigr| \bigl|\overset{^{\prime }}{F}_{\bar{f}}\bigr|}{%
\bigl|F_{f}\bigr|^{2}+\bigl|\overset{^{\prime }}{F}_{\bar{f}}\bigr|^{2}}\sin
\Delta mt\sin \bigl(2\phi _{M}-\phi -\phi ^{^{\prime }}\bigr) \cos \bigl(%
\delta _{f}-\overset{^{\prime }}{\delta }_{\bar{f}}\bigr)  \label{e6} \\
\mathcal{F}\left( t\right) &\equiv &\frac{\left[ \Gamma _{f}(t)+\bar{\Gamma}%
_{\bar{f}}\right] -\left[ \Gamma _{\bar{f}}(t)+\bar{\Gamma}_{f}\right] }{%
\left[ \Gamma _{f}(t)+\bar{\Gamma}_{\bar{f}}\right] +\left[ \Gamma _{\bar{f}%
}(t)+\bar{\Gamma}_{{f}}\right] }  \notag \\
&=&\frac{\bigl|F_{f}\bigr|^{2}-\bigl|\overset{^{\prime }}{F}_{\bar{f}}\bigr|%
^{2}}{\bigl|F_{f}\bigr|^{2}+\bigl|\overset{^{\prime }}{F}_{\bar{f}}\bigr|^{2}%
}\cos \Delta mt  \notag \\
&-&\frac{2\bigl|F_{f}\bigr| \bigl|\overset{^{\prime }}{F}_{\bar{f}}\bigr|}{%
\bigl|F_{f}\bigr|^{2}+\bigl|\overset{^{\prime }}{F}_{\bar{f}}\bigr|^{2}}\sin
\Delta mt\cos \left( 2\phi _{M}-\phi -\phi ^{^{\prime }}\right) \sin \bigl(%
\delta _{f}-\overset{^{\prime }}{\delta }_{\bar{f}}\bigr)  \label{e7}
\end{eqnarray}

We now apply the above formula to $B\rightarrow \pi D$ and $B_{s}\rightarrow
KD_{s}$ decays. For these decays, 
\begin{equation*}
\phi =0,\qquad \phi ^{\prime }=\gamma
\end{equation*}%
\begin{equation*}
\phi _{M}=%
\begin{cases}
-\beta , & \text{for $B^{0}$} \\ 
-\beta _{s}, & \text{for $B_{s}^{0}$}%
\end{cases}%
\end{equation*}%
\begin{eqnarray*}
A_{f} &=&\langle D^{-}\pi ^{+}\left\vert \mathcal{L_{W}}\right\vert
B^{0}\rangle =F_{f} \\
A_{\bar{f}}^{\prime } &=&\langle D^{+}\pi ^{-}\left\vert \mathcal{L_{W}}%
^{\prime }\right\vert B^{0}\rangle =e^{i\gamma }F_{\bar{f}}^{^{\prime }} \\
A_{f_{s}} &=&\langle K^{+}D_{s}^{-}\left\vert \mathcal{L_{W}}\right\vert
B_{s}^{0}\rangle =F_{f_{s}} \\
A_{\bar{f}_{s}}^{\prime } &=&\langle K^{-}D_{s}^{+}\left\vert \mathcal{L_{W}}%
^{\prime }\right\vert B_{s}^{0}\rangle =e^{i\gamma }F_{\bar{f}%
_{s}}^{^{\prime }}
\end{eqnarray*}

Note that the effective Lagrangians for decays $(q=d,s)$ are given by, 
\begin{subequations}
\begin{eqnarray}
&&\mathcal{L_{W}}=V_{cb}V_{uq}^{\ast }\left[ \bar{q}\gamma ^{\mu }\left(
1-\gamma _{5}\right) u\right] \left[ \bar{c}\gamma _{\mu }\left( 1-\gamma
_{5}\right) b\right]  \label{e8a} \\
&&\mathcal{L_{W}}^{\prime}=V_{ub}V_{cq}^{\ast }\left[ \bar{q}\gamma ^{\mu
}\left( 1-\gamma _{5}\right) c\right] \left[ \bar{u}\gamma _{\mu }\left(
1-\gamma _{5}\right) b\right]  \label{e8b}
\end{eqnarray}

respectively. In the Wolfenstein parametrization of $CKM$ matrix, 
\end{subequations}
\begin{equation}
\frac{\left| V_{ub}\right| \left| V_{cq}\right| }{\left| V_{cb}\right|
\left| V_{uq}\right| } =\lambda ^{2}\sqrt{\bar{\rho}^{2}+\bar{\eta}^{2}},
\qquad  \label{e9}
\end{equation}

Define, 
\begin{equation*}
r=\lambda ^{2}R_{b}\frac{\bigl| \overset{^{\prime }}{F}_{\bar{f}}\bigr|}{%
\bigl| F_{f}\bigr|}andr_{s}=R_{b}\frac{\bigl| \overset{^{\prime }}{F}_{\bar{f%
}_{s}}\bigr|}{\bigl| F_{f_{s}}\bigr|}
\end{equation*}

Thus, we get from Eqs. $\eqref{e6}$ and $\eqref{e7}$ for $B^{0}$ decays,
(replacing $\frac{\bigl| \overset{^{\prime }}{F}_{\bar{f}}\bigr|}{\bigl| %
F_{f}\bigr|}$ by $r$),

\begin{eqnarray}
\mathcal{A}\left( t\right) &=&-\frac{2r}{1+r^{2}}\sin \Delta m_{B}t\sin
\left( 2\beta +\gamma \right) \cos \left( \delta _{f}-\overset{^{\prime }}{%
\delta }_{\bar{f}}\right)  \notag \\
\mathcal{F}\left( t\right) &=&\frac{1-r^{2}}{1+r^{2}}\cos \Delta m_{B}t-%
\frac{2r}{1+r^{2}}\sin \Delta m_{B}t\cos \left( 2\beta +\gamma \right) \sin
\left( \delta _{f}-\overset{^{\prime }}{\delta }_{\bar{f}}\right)
\label{4.35b}
\end{eqnarray}

For the decays, 
\begin{eqnarray*}
\bar{B}_{s}^{0}\left( B_{s}^{0}\right) &\rightarrow &K^{-}D_{s}^{+}\left(
K^{+}D_{s}^{-}\right) \\
\bar{B}_{s}^{0}\left( B_{s}^{0}\right) &\rightarrow &K^{+}D_{s}^{-}\left(
K^{-}D_{s}^{+}\right)
\end{eqnarray*}

we get, 
\begin{eqnarray}
\mathcal{A}_{s}\left( t\right) &=& - \frac{2r_{s}}{1+r_{s}^{2}}\sin (\Delta
m_{B_{s}}t)\sin \left( 2\beta _{s} + \gamma \right) \cos \left( \delta
_{f_{s}}-\overset{^{\prime }}{\delta }_{\bar{f}_{s}}\right)  \notag \\
\mathcal{F}_{s}(t) &=&\frac{1-r_{s}^{2}}{1+r_{s}^{2}}\cos \Delta m_{B_{s}}t-%
\frac{2r_{s}}{1+r_{s}^{2}}\sin \Delta m_{B_{s}}t\cos \left( 2\beta _{s} +
\gamma \right) \sin \left( \delta _{f_{s}}-\overset{^{\prime }}{\delta }_{%
\bar{f}_{s}}\right)  \label{4.35}
\end{eqnarray}

We note that for time integrated $CP$-asymmetry, 
\begin{eqnarray}
\mathcal{A}_{s} &\equiv &\frac{\int_{0}^{\infty }\left[ \Gamma _{fs}\left(
t\right) -\bar{\Gamma}_{fs}\left( t\right) \right] dt}{\int_{0}^{\infty }%
\left[ \Gamma _{fs}\left( t\right) +\bar{\Gamma}_{fs}\left( t\right) \right]
dt}  \notag \\
&=&-\frac{2r}{1+r^{2}}\sin \left( 2\beta _{s} + \gamma \right) \frac{\Delta
m_{B_{s}}/\Gamma _{s}}{1+\left( \Delta m_{B_{s}}/\Gamma _{s}\right) ^{2}}%
\cos (\delta _{f_{s}}-\overset{^{\prime }}{\delta }_{\bar{f}_{s}})
\label{4.35a}
\end{eqnarray}

The $CP$--asymmetry $\mathcal{A}_{s}\left( t\right) $ or $\mathcal{A}_{s}$
involves two experimentally unknown parameters $\sin \left( 2\beta
_{s}-\gamma \right) $ and $\Delta m_{B_{s}}$. Both these parameters are of
importance in order to test the unitarity of $CKM$ matrix viz whether $CKM$
matrix is a sole source of $CP$--violation in the processes in which $CP$%
--violation has been observed.

Within the \textbf{case II}, we discuss $B$ decays into baryons and
antibaryons.

So far we have discussed the $CP$-violation in kaon and $B_{q}^{0}-\bar{B}%
_{q}^{0}$ systems. There is thus a need to study $CP$-violation outside
these systems.

The decays of $B(\bar{B})$ mesons to baryon-antibaryon pair $N_{1}$ $\bar{N}%
_{2}$ $(\bar{N}_{1}$ $N_{2})$ and subsequent decays of $N_{2},\bar{N}_{2}$
or $(N_{1},\bar{N}_{1})$ to a lighter hyperon (antihyperon) plus a meson
provide a means to study $CP$-odd observables as for example in the process, 
\begin{equation*}
e^{-}e^{+}\rightarrow B,\bar{B}\rightarrow N_{1}\bar{N}_{2}\rightarrow N_{1}%
\bar{N}_{2}^{\prime }\bar{\pi},\qquad \bar{N}_{1}N_{2}\rightarrow \bar{N}%
_{1}N_{2}^{\prime }\pi
\end{equation*}

The decay $B\rightarrow N_{1}\bar{N}_{2}(f)$ is described by the matrix
element, 
\begin{equation}
M_{f}=F_{q}e^{+i\phi }\left[ \bar{u}(\mathbf{p}_{1})(A_{f}+\gamma
_{5}B_{f})v(\mathbf{p}_{2})\right]  \label{q1}
\end{equation}%
where as $B\rightarrow \overline{N}_{1}N_{2}(\overline{f})$ is described by
the matrix elements 
\begin{equation*}
\overset{}{M^{\prime }}_{f}=\overset{}{F^{\prime }}_{q}e^{+i\phi ^{^{\prime
}}}\left[ \bar{u}(\mathbf{p}_{2})(\overset{}{A^{\prime }}_{\overline{f}%
}+\gamma _{5}\overset{}{B^{\prime }}_{\overline{f}})v(\mathbf{p}_{1})\right]
\end{equation*}

where $F_{q}$ is a constant containing CKM factor, $\phi $ is the weak
phase. The amplitude $A_{f}$ and $B_{f}$ are in general complex in the sense
that they incorporate the final state phases $\delta _{p}^{f}$ and $%
\delta_{s}^{f}$ and they may also contain weak phases $\phi_{s}$ and $%
\phi_{p}$ Note that $A_{f}$ is the parity violating amplitude ($p$-wave)
whereas $B_{f}$ is parity conserving amplitude ($s$-wave). The $CPT$
invariance gives the matrix elements for the decay $\bar{B}\rightarrow \bar{%
N }_{1}N_{2}(\bar{f}):$

\begin{equation}
\bar{M}_{\bar{f}}=F_{q}e^{-i\phi }\left[ \bar{u}(\mathbf{p}%
_{2})(-A_{f}^{\ast }e^{2i\delta _{p}^{f}}+\gamma _{5}B_{f}^{\ast
}e^{2i\delta _{s}^{f}})v(\mathbf{p}_{1})\right]  \label{q2}
\end{equation}

if the decays are described by a single matrix element $M_{f}$. If $%
\phi_{s}=0=\phi_{p}$ then $CPT$ and $CP$ invariance give the same
predictions viz 
\begin{equation}
\bar{\Gamma}_{\bar{f}}=\Gamma _{f}, \qquad \bar{\alpha}_{\bar{f}}=-\alpha
_{f}, \qquad \bar{\beta}_{\bar{f}}=-\beta _{f}, \qquad \bar{\gamma}_{\bar{f}
}=\gamma _{f}  \label{q3}
\end{equation}

The decay width for the mode $B\rightarrow N_{1}\bar{N}_{2}(f)$ is given by, 
\begin{eqnarray}
\Gamma _{f} &=&\frac{m_{1}m_{2}}{2\pi m_{B}^{2}}\left| \mathbf{p}\right|
\left| M_{f}\right| ^{2}  \notag \\
&=&\frac{F_{q}^{2}}{2\pi m_{B}^{2}}\left| \mathbf{p}\right| \left[
(p_{1}\cdot p_{2}-m_{1}m_{2})\left| A_{f}\right| ^{2}+(p_{1}\cdot
p_{2}+m_{1}m_{2})\left| B_{f}\right| ^{2}\right]  \label{q4}
\end{eqnarray}

In order to take into account the polarization of $N_{1}$ and $\bar{N}_{2},$
we give the general expression for $\left| M_{f}\right| ^{2}$, 
\begin{eqnarray}
\left| M_{f}\right| ^{2} &=&\frac{F_{q}^{2}}{16m_{1}m_{2}}Tr\left[ 
\begin{array}{c}
(\not{p}_{1}+m_{1})(1+\gamma _{5}\gamma \cdot s_{1})(A_{f}+\gamma
_{5}B_{f})( \not{p}_{2}-m_{2}) \\ 
\times (1+\gamma _{5}\gamma \cdot s_{2})(A_{f}^{\ast }-\gamma
_{5}B_{f}^{\ast })%
\end{array}
\right]  \label{q5}
\end{eqnarray}

where $s_{1}^{\mu}, s_{2}^{\mu}$ are polarization vectors of $N_{1}$ and $%
\bar{N}_{2}$ respectively $(p_{1}.s_{1}=0, \quad p_{2}.s_{2}=0, \quad
s_{1}^{2}=-1=s_{2}^2)$.

In the rest frame of $B$, we get, 
\begin{eqnarray}
\left\vert M_{f}\right\vert ^{2} &=&F_{q}^{2}\frac{2E_{1}E_{2}}{4m_{1}m_{2}}%
\left[ \left\vert a_{s}^{f}\right\vert ^{2}+\left\vert a_{p}^{f}\right\vert
^{2}\right]  \notag \\
&&\left\{ 
\begin{array}{c}
1+\alpha _{f}\left( \frac{m_{1}}{E_{1}}\mathbf{n}\cdot \mathbf{s}_{1}-\frac{%
m_{2}}{E_{2}}\mathbf{n}\cdot \mathbf{s}_{2}\right) \\ 
+\beta _{f}\mathbf{n}\cdot (\mathbf{s}_{1}\times \mathbf{s}_{2})+\gamma _{f}%
\left[ (\mathbf{n}\cdot \mathbf{s}_{1})(\mathbf{n}\cdot \mathbf{s}_{2})-%
\mathbf{s}_{1}\cdot \mathbf{s}_{2}\right] \\ 
-\frac{m_{1}m_{2}}{E_{1}E_{2}}(\mathbf{n}\cdot \mathbf{s}_{1})(\mathbf{n}%
\cdot \mathbf{s}_{2})%
\end{array}%
\right\}  \label{q6}
\end{eqnarray}

where, 
\begin{eqnarray}
a_{s} &=&\sqrt{\frac{p_{1}\cdot p_{2}+m_{1}m_{2}}{2E_{1}E_{2}}}B, \quad
a_{p}=-\sqrt{\frac{p_{1}\cdot p_{2}-m_{1}m_{2}}{2E_{1}E_{2}}}A  \label{q7} \\
\alpha _{f} &=&\frac{2S_{f}P_{f}\cos (\delta _{s}^{f}-\delta _{p}^{f})}{
S_{f}^{2}+P_{f}^{2}}, \quad \beta _{f}=\frac{2S_{f}P_{f}\sin (\delta
_{s}^{f}-\delta _{p}^{f})}{S_{f}^{2}+P_{f}^{2}}  \notag \\
\gamma _{f} &=&\frac{S_{f}^{2}-P_{f}^{2}}{S_{f}^{2}+P_{f}^{2}}, \quad
a_{s}=S_{f}e^{i\delta _{s}^{f}}, \quad a_{p}^{f}=P_{f}e^{i\delta _{p}^{f}}
\label{q8}
\end{eqnarray}

In the rest frame of $B$, due to spin conservation, 
\begin{equation}
\left( \lambda _{1}\equiv \frac{E_{1}}{m_{1}}\mathbf{n}\cdot \mathbf{s}%
_{1}\right) =\left( \lambda _{2}\equiv \frac{-E_{2}}{m_{2}}\mathbf{n}\cdot 
\mathbf{s}_{2}\right) =\pm 1  \label{q9}
\end{equation}

Thus, invariants multiplying $\beta _{f}$ and $\gamma _{f}$ vanish. Hence we
have, 
\begin{eqnarray}
\left| M_{f}\right| ^{2} &=&\left( \frac{2E_{1}E_{2}}{m_{1}m_{2}}\right)
F_{q}^{2}(S_{f}^{2}+P_{f}^{2})\left[ (1+\lambda _{1}\lambda _{2})+\alpha
_{f}(\lambda _{1}+\lambda _{2})\right]  \label{q10} \\
\Gamma _{f} &=&\Gamma _{f}^{++}+\Gamma _{f}^{--}=\frac{2E_{1}E_{2}}{2\pi
m_{B}^{2}}\left| \vec{p}\right| F_{q}^{2}\left[ S_{f}^{2}+P_{f}^{2}\right] = 
\bar{\Gamma}_{\bar{f}}  \label{q11} \\
\Delta \Gamma _{f} &=&\frac{\Gamma _{f}^{++}-\Gamma _{f}^{--}}{\Gamma
_{f}^{++}+\Gamma _{f}^{--}}=\alpha _{f}, \qquad \Delta \bar{\Gamma}_{\bar{f}
}=\bar{\alpha}_{\bar{f}}=-\alpha _{f}  \label{q12}
\end{eqnarray}

Eqs. (\ref{q11}) and (\ref{q12}) follow from $CP$ invariance. It will be of
interest to test these equations.

Now $B_{q}^{0},$ $\bar{B}_{q}^{0}$ annihilate into baryon-antibaryon pair $%
N_{1}\bar{N}_{2}$ through $W$-exchange as depicted in Figs (5a) and (5b). $%
B^{-}\rightarrow N_{1}\bar{N}_{2}$ through annihilation diagram is shown in
Fig (6). It is clear from Fig (5a) and (5b), that we have the same final
state configuration for $B_{q}^{0},$ $\bar{B}_{q}^{0}\rightarrow N_{1}\bar{N}%
_{2}.$ Thus, one would expect, 
\begin{eqnarray}
S_{\bar{f}}^{^{\prime }} &=&S_{f},\qquad P_{\bar{f}}^{^{\prime }}=P_{f} 
\notag \\
\overset{^{\prime }}{\delta }_{s}^{\bar{f}} &=&\delta _{s}^{f},\qquad 
\overset{^{\prime }}{\delta }_{p}^{\bar{f}}=\delta _{p}^{f}  \label{24}
\end{eqnarray}

Hence we have, 
\begin{eqnarray}
\Gamma _{\bar{f}}^{\prime } &=&\bar{\Gamma}_{f}^{\prime }=r^{2}\Gamma _{f};\
\ \ \ r^{2}=\frac{\left\vert F_{q}^{\prime }\right\vert ^{2}}{\left\vert
F_{q}\right\vert ^{2}}  \label{25} \\
\bar{\alpha}_{f}^{\prime } &=&-\alpha _{\bar{f}}^{\prime }=\alpha _{f}=-\bar{%
\alpha}_{\bar{f}}  \label{26}
\end{eqnarray}

Above predictions can be tested in future experiments on baryon decay modes
of $B$-mesons. In particular $\bar{\alpha}_{f}^{\prime }=\alpha _{f}$ would
give direct confirmation of Eqs. (\ref{24}).

For the time dependent baryon decay modes of $B_{q}^{0}-\bar{B}_{q}^{0}$, we
have: $\left( \phi =\gamma ,\ \phi ^{\prime }=0\right) $ 
\begin{eqnarray}
\mathcal{A(}t) &=&\mathcal{A}^{++}(t)+\mathcal{A}^{--}(t)=\frac{2r\sin
\Delta mt\sin (2\phi _{M}-\gamma )}{1+r^{2}}  \label{46} \\
\Delta \mathcal{A}(t) &=&\mathcal{A}^{++}(t)-\mathcal{A}^{--}(t)=0
\label{47} \\
\mathcal{F}(t) &=&\mathcal{F}^{++}(t)+\mathcal{F}^{--}(t)=\frac{1-r^{2}}{%
1+r^{2}}\cos \Delta mt  \label{48} \\
\Delta \mathcal{F}(t) &=&\mathcal{F}^{++}(t)-\mathcal{F}^{--}(t)=\frac{%
1-r^{2}}{2(1+r^{2})}(\alpha _{f}+\bar{\alpha}_{\bar{f}})\cos \Delta mt 
\notag \\
&&-\frac{4r\sin \Delta mt\sin (2\phi _{M}-\gamma )S_{f}P_{f}}{%
(1+r^{2})(S_{f}^{2}+P_{f}^{2})}  \label{49}
\end{eqnarray}%
where we have used Eqs. (\ref{26}). For $B_{d}^{0},$ $r=-\lambda ^{2}\sqrt{%
\bar{\rho}^{2}+\bar{\eta}^{2}}\approx -(0.02\pm 0.006)$ [4], $\phi
_{M}=-\beta ;$ for $B_{s}^{0},$ $r=-\sqrt{\bar{\rho}^{2}+\bar{\eta}^{2}}%
\approx -(0.40\pm 0.13)$, $\phi _{M}=-\beta _{s}$.

Eq.(\ref{46}) gives a means to determine the weak phase $2\beta +\gamma $ or 
$\gamma $ in the baryon decay modes of $B_{d}^{0}$ and $B_{s}^{0}$
respectively. Non-zero $\cos \Delta mt$ term in $\Delta \mathcal{F}(t)$
would give clear indication of $CP$ violation especially for baryon decay
modes of $B_{d}^{0},$ for which \thinspace $r^{2}\leq 1,$ so that $\frac{%
1-r^{2}}{1+r^{2}}\approx 1$. It may be noted that the time-dependent
asymmetries arises because there are two independent amplitudes for the
decays $B_{q}^{0}\rightarrow N_{1}\overline{N}_{2},$ $\overline{N}%
_{1}N_{2}:M_{f}\overset{}{,\ M^{\prime }}_{\overline{f}}.$

The baryon decay modes of $B$-mesons not only provide a means to test
prediction of $CP$ asymmetry viz $\alpha _{f}+\bar{\alpha}_{\bar{f}}=0$ for
charmed baryons (discussed above) but also to test the $CP$-asymmetry in
hyperon (antihyperon) decays viz absence of $CP$-odd observables $\Delta
\Gamma ,\Delta \alpha ,\Delta \beta $ discussed in [8]. Consider for example
the decays, 
\begin{equation*}
B_{q}^{0}\rightarrow p\bar{\Lambda}_{c}^{-}\rightarrow p\bar{p}K^{0}(p\bar{%
\Lambda}\pi ^{-}\rightarrow p\overline{p}\pi ^{+}\pi ^{-}),
\end{equation*}%
\begin{equation*}
\bar{B}_{q}^{0}\rightarrow \overline{p}\Lambda _{c}^{+}\rightarrow \bar{p}p%
\bar{K}^{0}(\overline{p}\Lambda \pi ^{+}\rightarrow \bar{p}p\overline{b}\pi
^{-}\pi ^{+})
\end{equation*}

By analyzing the final state $\bar{p}p\bar{K}^{0},p\bar{p}K^{0},$ one may
test $\alpha _{f}=-\bar{\alpha}_{\bar{f}}$ for the charmed hyperon. We note
that for $\Lambda _{c}^{+},$ $c\tau =59.9\mu $m, whereas $c\tau =7.8$cm for $%
\Lambda -$hyperon, so that the decays of $\Lambda _{c}^{+}$ and $\Lambda $
would not interfere with each other. By analysing the final state $\bar{p}%
p\pi ^{-}\pi ^{+}$ and $p\bar{p}\pi ^{+}\pi ^{-},$ one may check $CP$%
--violation for hyperon decays. One may also note that for $(B_{d}^{0},\bar{B%
}_{d}^{0})$ complex, the competing channels viz $B_{d}^{0}\rightarrow \bar{p}%
\Lambda _{c}^{+},$ $\bar{B}_{d}^{0}\rightarrow p\bar{\Lambda}_{c}^{-}$ are
doubly Cabibbo suppressed by $r^{2}=\lambda ^{2}\left( \bar{\rho}^{2}+\bar{%
\eta}^{2}\right) $ unlike $(B_{s}^{0}-\bar{B}_{s}^{0})$ complex where the
competing channels are suppressed by a factor of $\left( \bar{\rho}^{2}+\bar{%
\eta}^{2}\right) $. Hence $B_{d}^{0}($ $\bar{B}_{d}^{0})$ decays are more
suitable for this type of analysis. Other decays of interest are, 
\begin{eqnarray*}
B^{-} &\rightarrow &\Lambda \bar{\Lambda}_{c}^{-}\rightarrow \Lambda \bar{%
\Lambda}\pi ^{-}\rightarrow p\pi ^{-}\bar{p}\pi ^{+}\pi ^{-} \\
B^{+} &\rightarrow &\bar{\Lambda}\Lambda _{c}^{+}\rightarrow \bar{\Lambda}%
\Lambda \pi ^{+}\rightarrow \bar{p}\pi ^{+}p\pi ^{-}\pi ^{+} \\
B_{c}^{-} &\rightarrow &\bar{p}\Lambda \rightarrow \bar{p}p\pi ^{-} \\
B_{c}^{+} &\rightarrow &p\bar{\Lambda}\rightarrow p\bar{p}\pi ^{+}
\end{eqnarray*}

The non-leptonic hyperon (antihyperon) decays $N\rightarrow N^{\prime }\pi (%
\bar{N}\rightarrow \bar{N}^{\prime }\bar{\pi})$ are related to each other by 
$CPT$, 
\begin{eqnarray*}
a_{l}(I) &=&\left\langle f_{lI}^{out}\left| H_{W}\right| N\right\rangle
=\eta _{f}e^{2i\delta _{l}(I)}\left\langle \bar{f}_{lI}^{out}\left|
H_{W}\right| \bar{N}\right\rangle \\
&=&\eta _{f}e^{2i\delta _{l}(I)}\bar{a}_{l}^{\ast }(I)
\end{eqnarray*}

Hence, 
\begin{equation*}
\bar{a}_{l}(I)=\eta _{f}e^{2i\delta _{l}(I)}\bar{a}_{l}^{\ast
}(I)=(-1)^{l+1}e^{i\delta _{l}(I)}e^{-i\phi }\left| a_{l}\right|
\end{equation*}

where we have selected the phase $\eta _{f}=(-1)^{l+1}$. Here $I$ is the
isospin of the final state and $\phi $ is the weak phase. Thus necessary
condition for non-zero $CP$ odd observables is that the weak phase for each
partial wave amplitude should be different. For instance for the decays $%
B^{0}(\bar{B}^{0})\rightarrow p\bar{\Lambda}_{c}^{-}(\bar{p}\Lambda
_{c}^{+}) $ we have, 
\begin{eqnarray*}
\delta \Gamma &=&0 \\
\delta \alpha _{f} &=&-\tan \left( \delta _{s}-\delta _{p}\right) \tan
\left( \phi _{s}-\phi _{p}\right) \\
&\approx &-\tan \left( \delta _{s}-\delta _{p}\right) \sin \left( \phi
_{s}-\phi _{p}\right)
\end{eqnarray*}

\bigskip

\textbf{Case III:} 

Here $A_{f}\neq A_{\bar{f}}$. 
\begin{eqnarray*}
A_{f} &=&\langle f\left| \mathcal{L_{W}}\right| B^{0}\rangle =\left[
e^{i\phi _{1}}F_{1f}+e^{i\phi _{2}}F_{2f}\right] \\
A_{\bar{f}} &=&\langle \bar{f}\left| \mathcal{L_{W}}\right| B^{0}\rangle =%
\left[ e^{i\phi _{1}}F_{1\bar{f}}+e^{i\phi _{2}}F_{2\bar{f}}\right]
\end{eqnarray*}

Examples: 
\begin{equation*}
B^{0} \rightarrow \rho ^{-}\pi ^{+}(f):\text{ }A_{f} \qquad B^{0}
\rightarrow \rho ^{+}\pi ^{-}(\bar{f}):A_{\bar{f}}
\end{equation*}
\begin{equation*}
B_{s}^{0} \rightarrow K^{\ast -}K^{+} \qquad B_{s}^{0} \rightarrow K^{\ast
+}K^{-}
\end{equation*}

$CPT$ gives, 
\begin{equation*}
\bar{A}_{\bar{f},f}=\sum_{i}[e^{-i\phi_{i}}e^{2i\delta^{i}_{f,\bar{f}}}
F^{\ast}_{i f \bar{f}}]
\end{equation*}

Subtracting and adding Eqs. (\ref{e2}) and (\ref{e1}), we get,

We now discuss the decays listed in case (ii) where $A_{f}\neq A_{\bar{f}}$.
Subtracting and adding Eqs. $(\ref{e2})$ and $(\ref{e1})$, we get, 
\begin{align}
\frac{\Gamma _{f}(t)-\bar{\Gamma}_{f}(t)}{\Gamma _{f}(t)+\bar{\Gamma}_{f}(t)}%
=& C_{f}\cos \Delta mt+S_{f}\sin \Delta mt  \notag \\
=& (C-\Delta C)\cos \Delta mt+(S-\Delta S)\sin \Delta mt  \label{ccc2} \\
\frac{\Gamma _{\bar{f}}(t)-\bar{\Gamma}_{\bar{f}}(t)}{\Gamma _{\bar{f}}(t)+%
\bar{\Gamma}_{\bar{f}}(t)}=& C_{\bar{f}}\cos \Delta mt+S_{\bar{f}}\sin
\Delta mt  \notag \\
=& (C+\Delta C)\cos \Delta mt+(S+\Delta S)\sin \Delta mt  \label{ccc3}
\end{align}

where 
\begin{align}
C_{\bar{f},f} &= (C\pm\Delta C)  \notag \\
&= \frac{\bigl|A_{\bar{f},f}\bigr|^{2} - \bigl|\bar{A}_{\bar{f},f}\bigr|^{2}%
}{\bigl|A_{\bar{f},f}\bigr|^{2} + \bigl|\bar{A}_{\bar{f},f}\bigr|^{2}} 
\notag \\
&= \frac{\Gamma_{\bar{f},f}-\bar{\Gamma}_{\bar{f},f}}{\Gamma_{\bar{f},f} + 
\bar{\Gamma}_{\bar{f},f}}  \notag \\
&= \frac{R_{\bar{f}, f} (1 - A_{CP}^{\bar{f}, f}) - R_{\bar{f}, f} (1 +
A_{CP}^{\bar{f}, f})}{\Gamma (1 \pm A_{CP})}  \label{ccc4} \\
S_{\bar{f},f} &= (S \pm \Delta S) \\
&= \frac{2\text{Im} [e^{2i\phi_{M}}A^{\ast}_{\bar{f},f}\bar{A}_{\bar{f},f}]}{%
\Gamma_{\bar{f},f} + \bar{\Gamma}_{\bar{f},f}}  \label{ccc5} \\
A_{CP}^{\bar{f}} &= \frac{\bar{\Gamma}_{f}-\Gamma _{\bar{f}}}{\Gamma _{\bar{f%
}}+\bar{\Gamma}_{f}}  \notag \\
A_{CP}^{f} &= \frac{\bar{\Gamma}_{\bar{f}} - \Gamma _{f}}{\Gamma _{f} + \bar{%
\Gamma}_{\bar{f}}}  \label{ccc6} \\
A_{CP}&= \frac{(\Gamma_{\bar{f}} + \bar{\Gamma}_{\bar{f}}) - (\bar{\Gamma_{f}%
} + \Gamma_{f})}{(\Gamma_{\bar{f}} - \bar{\Gamma}_{\bar{f}}) - (\bar{%
\Gamma_{f}} + \Gamma_{f})} \\
&= \frac{R_{f}A^{f}_{CP}-R_{\bar{f}}A^{\bar{f}}_{CP}}{\Gamma}  \label{ccc7}
\end{align}

where 
\begin{align}
R_{f} &= \frac{1}{2}(\Gamma_{f} + \bar{\Gamma}_{\bar{f}}), \qquad R_{\bar{f}%
}=\frac{1}{2}(\Gamma_{\bar{f}} + \bar{\Gamma}_{f})  \notag \\
\Gamma &= R_{f}+R_{\bar{f}}  \label{ccc8}
\end{align}

The following relations are also useful which can be easily derived from
above equations 
\begin{align}
\frac{R_{\bar{f},f}}{R_{f}+R_{\bar{f}}}& =\frac{1}{2}[(1\pm \Delta C)\pm
A_{CP}C]  \label{ccc9} \\
\frac{R_{\bar{f}}-R_{f}}{R_{f}+R_{\bar{f}}}& =[\Delta C+A_{CP}C]
\label{ccc10} \\
\frac{R_{\bar{f}}A_{CP}^{\bar{f}}+R_{f}A_{CP}^{f}}{R_{f}+R_{\bar{f}}}&
=[C+A_{CP}\Delta C]  \label{ccc11}
\end{align}

For these decays, the decay amplitudes can be written in terms of tree
amplitude $e^{i\phi_{T}}T_{f}$ and the penguin amplitude $e^{i\phi_{P}}P_{f}$%
: 
\begin{align}
A_{f} &= e^{i\phi_{T}}e^{i\delta_{f}^{T}}\bigl|T_{f}\bigr| \lbrack 1 +
r_{f}e^{i(\phi_{P}-\phi_{T})}e^{i\delta_{f}}]  \notag \\
A_{\bar{f}} &= e^{i\phi_{T}}e^{i\delta_{\bar{f}}^{T}}\bigl|T_{\bar{f}}\bigr| %
\lbrack 1 + r_{\bar{f}} e^{i(\phi_{P}-\phi_{T})}e^{i\delta_{\bar{f}}}]
\label{ccc12}
\end{align}

where $r_{f, \bar{f}} = \frac{\bigl|P_{f,\bar{f}}\bigr|}{\bigl|T_{f,\bar{f}}%
\bigr|}, \quad \delta_{f, \bar{f}} = \delta^{P}_{f, \bar{f}} - \delta^{T}_{f,%
\bar{f}} $. 
\begin{align}
\bar{A}_{\bar{f}} &= e^{-i\phi_{T}}e^{i\delta_{f}^{T}}\bigl|T_{f}\bigr| %
\lbrack 1 + r_{f}e^{-i(\phi_{P}-\phi_{T})}e^{i\delta_{f}}]  \notag \\
\bar{A}_{f} &= e^{-i\phi_{T}}e^{i\delta_{\bar{f}}^{T}}\bigl|T_{\bar{f}}%
\bigr| \lbrack 1 + r_{\bar{f}} e^{-i(\phi_{P}-\phi_{T})}e^{i\delta_{\bar{f}%
}}]  \label{ccc13}
\end{align}

\begin{equation}
\text{For} B^{0} \rightarrow \rho^{-} \pi^{+}: A_{f}; \qquad B^{0}
\rightarrow \rho^{+} \pi^{-}: A_{\bar{f}}; \quad \phi_{T} = \gamma, \phi_{P}
= -\beta  \label{ccc14}
\end{equation}

\begin{equation}
\text{For} B^{0} \rightarrow D^{\ast -}D^{+}: A^{D}_{f}; \qquad B^{0}
\rightarrow D^{\ast +}D^{-}: A^{D}_{\bar{f}}; \quad \phi_{T} = 0, \phi_{P} =
-\beta  \label{ccc15}
\end{equation}

Hence for $B^{0} \rightarrow \rho^{-} \pi^{+}, B^{0} \rightarrow \rho^{+}
\pi^{-}$, we have 
\begin{align}
A_{f} &= \bigl|T_{f}\bigr| e^{-i\gamma}e^{i\delta^{T}_{f}} [1 -
r_{f}e^{i(\alpha + \delta_{f})}]  \notag \\
A_{\bar{f}} &= \bigl|T_{\bar{f}}\bigr| e^{-i\gamma}e^{i\delta^{T}_{\bar{f}}}
[1 - r_{\bar{f}}e^{i(\alpha + \delta_{\bar{f}})}]  \label{ccc16} \\
\text{where} \qquad r_{f, \bar{f}} &= \frac{|V_{tb}| |V_{td}|}{|V_{ub}|
|V_{ud}|} \frac{\bigl|P_{f,\bar{f}} \bigr|}{\bigl|T_{f,\bar{f}} \bigr|} = 
\frac{R_{t}}{R_{b}} \frac{\bigl|P_{f,\bar{f}} \bigr|}{\bigl|T_{f,\bar{f}} %
\bigr|}  \label{ccc17}
\end{align}

and for $\text{B}^{0}\rightarrow D^{*-}D^{+}$, $\text{B}^{0}\rightarrow
D^{*+}D^{-}$, we have 
\begin{align}
A_{f}^{D}& =\bigl|T_{f}^{D}\bigr| e^{i\delta
_{f}^{TD}}[1-r_{f}^{D}e^{i(-\beta +\delta _{f}^{D})}]  \notag \\
A_{\bar{f}}^{D}& =\bigl|T_{\bar{f}}^{D}\bigr| e^{i\delta _{\bar{f}%
}^{TD}}[1-r_{\bar{f}}^{D}e^{i(-\beta +\delta _{\bar{f}}^{D})}]  \label{ccc18}
\\
\text{where}\qquad r_{f,\bar{f}}& =R_{t}\frac{\bigl|P_{f,\bar{f}}^{D}\bigr|}{%
\bigl|T_{f,\bar{f}}^{D}\bigr|}  \notag
\end{align}

\section{Final State Strong Phases}

As we have seen the CP asymmetries in the hadronic decays of B and K mesons
involve strong final state phases. Thus strong interactions in these decays
play a crucial role. The short distance strong interactions effects at the
quark level are taken care of by perturbative QCD in terms of Wilson
coefficients. The CKM matrix which connects the weak eigenstates will mass
eigenstates is another aspect of strong interactions at quark level. In the
case of semi leptonic decays, the long distance strong interaction effects
manifest themselves in the form factors of final states after hadronization.
Likewise the strong interaction final state phases are long distance
effects. These phase shifts essentially arise in terms of S-matrix which
changes an 'in' state into an 'out' state viz. 
\begin{equation}
|f\rangle _{out}=S|f\rangle _{in}=e^{2i\delta _{f}}|f\rangle _{in}
\label{4.66}
\end{equation}

In fact, the CPT invariance of weak interaction Lagrangian gives for the
weak decay $B(\bar{B})\rightarrow f(\bar{f})$ 
\begin{equation}
\bar{A}_{\bar{f}}\equiv _{out}\langle \bar{f}|\mathcal{L}_{w}|\bar{B}%
\rangle=\eta_{f}e^{2i\delta_{f}}A_{f}{\ast}  \label{4.67}
\end{equation}

It is difficult to reliably estimate the final state strong phase shifts. It
involves the hadronic dynamics. However, using isospin, C-invariance of
S-matrix and unitarity of S-matrix, we can relate these phases. In this
regard, the decays $B^{0}\rightarrow f,\bar{f}$ described by two independent
single amplitudes $A_{f}$ and $A_{\bar{f}}^{\prime }$ discussed in section 4
case (ii) and the decays described by the weak amplitudes $A_{f}\ne A_{\bar{f%
}}$, described in section case (iii) are of interest

The invariance of S-matrix viz. $S_{\bar{f}}=S_{f}$ would imply 
\begin{equation*}
\delta _{f}=\delta _{\bar{f}}^{\prime },\qquad \delta _{1f}=\delta _{1\bar{f}%
},\qquad \delta _{2f}=\delta _{2\bar{f}}
\end{equation*}
In the above decays, b is converted into $b\rightarrow c(u)+\bar{u}+d$. In
particular, for the tree graph, the configuration is such that $\bar{u}$ and
d essentially go together into color singlet states will the third quark
c(u) recoiling; there is a significant probability that system will
hadronize as a two body final state. Thus at least for the tree amplitude $%
\delta _{f}^{T}$ should be equal to $\delta _{\bar{f}}^{T}$. To proceed
further, we use the unitarity of S-matrix to relate the final state strong
phases. The time reversal invariance gives 
\begin{equation}
F_{f}=_{out}\langle f|\mathcal{L}_{W}|B\rangle =_{in}\langle f|\mathcal{L}%
_{W}|B\rangle ^{*}  \label{4.68}
\end{equation}
where $\mathcal{L}_{W}$ is the weak interaction Lagrangian without the CKM
factor such as $V_{ud}^{*}V_{ub}$. From Eq. $\eqref{4.68}$, we have 
\begin{align}
F_{f}^{*}=& _{out}\langle f|S^{\dagger }\mathcal{L}_{W}|B\rangle  \notag \\
=& \sum_{n}S_{nf}^{*}F_{n}  \label{4.69}
\end{align}

It is understood that the unitarity equation which follows from time
reversal invaraince holds for each amplitude with the same weak phase. Above
equation can be written in two equivalent forms:

\begin{enumerate}
\item Exclusive version of Unitarity \newline

Writing 
\begin{equation}
S_{nf}=\delta _{nf}+iM_{nf}  \label{4.70}
\end{equation}

we get from Eq. $\eqref{4.69}$, 
\begin{equation}
ImF_{f}=\sum_{n}M_{nf}^{*}F_{n}  \label{4.71}
\end{equation}

where $M_{nf}$ is the scattering amplitude for $f\rightarrow n$. In this
version, the sum is over all allowed exclusive channels. This version is
more suitable in a situation where a single exclusive channel is dominant
one. To get the final result, one uses the dispersion relation.

\item Inclusive version of Unitarity \newline
This version is more suitable for our analysis. For this case, we write Eq. $%
\eqref{4.69}$ in the form 
\begin{equation}
F_{f}^{*}-S_{ff}^{*}F_{f}=\sum_{n\neq f}S_{nf}^{*}F_{n}  \label{4.72}
\end{equation}
\end{enumerate}

Parametrizing S-matrix as $S_{ff} \equiv S = \eta e^{2i\Delta}$, we get
after taking the absolute square of both sides of Eq. $\eqref{4.72}$ 
\begin{equation}
|F_{f}|^{2} [(q + \eta^{2}) - 2\eta \cos 2(\delta_{f} - \Delta)] = \sum_{n,
n^{\prime} \neq f} F_{n}S_{nf}^{\ast}F^{\ast}_{n^{\prime}}S_{n^{\prime}f}
\label{4.73}
\end{equation}

The above equation is an exact equation. In the random phase approximation,
we can put 
\begin{align}
\sum_{n^{\prime}, n \neq f} F_{n}S_{nf}^{\ast}F_{n^{\prime}}S_{n^{\prime}f}
=& \sum_{n \neq f} |F_{n}|^{2}|S_{nf}|^{2}  \notag \\
=& \bar{|F_{n}|^{2}} (1-\eta^{2})  \label{4.74}
\end{align}

We note that in a single channel description: 
\begin{equation*}
(Flux)_{in} - (Flux)_{out} = 1 - |\eta e^{2i\Delta}|^{2} = 1 - \eta^{2} = 
\text{Absorption}
\end{equation*}
The absorption takes care of all the inelastic channels. \newline
Similarly for the amplitude $F_{\bar{f}}$, we have 
\begin{equation}
F_{\bar{f}}^{\ast} - S^{\ast}_{\bar{f}\bar{f}}F_{\bar{f}} = \sum_{\bar{n}
\neq \bar{f}} S^{\ast}_{\bar{n}\bar{f}}F_{\bar{n}}  \label{4.75}
\end{equation}

The C-invariance of S-matrix gives: 
\begin{align}
S_{fn} =& \langle f|S|n\rangle = \langle f|C^{-1}CSC^{-1}C|n\rangle  \notag
\\
=& \langle \bar{f}|S|\bar{n}\rangle = S_{\bar{f}\bar{n}}  \label{4.76}
\end{align}

Thus in particular C-invariance of S-matrix gives 
\begin{equation}
S_{\bar{f}\bar{f}} = S_{ff} = \eta e^{2i\Delta}  \label{12}
\end{equation}

Hence from Eq. $\eqref{4.73}$, using Eqs. ($\ref{4.74}-\ref{12}$), we get 
\begin{equation}
\frac{1}{1-\eta ^{2}}[(1+\eta ^{2})-2\eta \cos 2(\delta _{f,\bar{f}}-\Delta
)]=\rho ^{2},\bar{\rho}^{2}  \label{4.78}
\end{equation}

where 
\begin{equation}
\rho^{2} = \frac{\overline{\bigl|F_{n}\bigr|}^{2}}{\bigl|F_{f}\bigr|^{2}},
\qquad \bar{\rho}^{2} = \frac{\overline{\bigl|F_{\bar{n}}\bigr|}^{2}}{\bigl|%
F_{\bar{f}}\bigr|^{2}}  \label{4.79}
\end{equation}

It is convenient to write Eq. $\eqref{4.78}$ in the form 
\begin{align}
\sin ^{2}(\delta _{f,\bar{f}}-\Delta )& =\frac{1-\eta ^{2}}{4\eta }\left[
\rho ^{2},\bar{\rho}^{2}-\frac{1-\eta }{1+\eta }\right]  \label{aa1} \\
0& \leq (\delta _{f,\bar{f}}-\Delta )\leq \theta  \label{aa2} \\
-\theta & \leq (\delta _{f,\bar{f}}-\Delta )\leq 0  \label{aa3}
\end{align}

where $\theta =\sin ^{-1}\sqrt{\frac{1-\eta }{2}}$.

The strong interaction parameters $\Delta \ $and$\ \eta \ $can be determond
by strong interaction dynamics. Using $SU\left( 2\right) $, C-invarience of
strong interactions and Regge pole phenomonology, the scattering aplitude $%
M\left( s,t\right) $ for two particle final state can be calculated.(For
details see ref. [12]). The s-wave scattering amplitude $f$ for the decay
modes $\pi ^{+}D^{-}\left( \pi ^{-}D^{+}\right) ,\ K^{+}\pi ^{-},\ \pi
^{+}\pi ^{-}$ which are s-wave decay modes of $B^{0}$ is given by%
\begin{equation*}
f\left( s\right) =\frac{1}{16\pi s}\int_{-s}^{0}M\left( s^{\prime }t\right)
dt
\end{equation*}%
where 
\begin{equation*}
t\approx -\frac{1}{2}s\left( 1-\cos \theta \right)
\end{equation*}%
Using the relation $S=\eta e^{2i\Delta }=1+2if$, the phase shift $\Delta ,$
the parameter $\eta $ and the phase angle $\theta $ can be determind. One
gets $\left( s=m_{B}^{2}\right) $%
\begin{eqnarray*}
\pi ^{+}D^{-}\left( \pi ^{-}D^{+}\right) &:&\ \Delta \approx -7^{o},\ \eta
\approx 0.62,\ \rho _{\min }\approx 0.23,\ \theta \approx 26^{o} \\
K^{+}\pi ^{-}\ or\ K^{0}\pi ^{+} &:&\ \Delta \approx -9^{o},\ \eta \approx
0.52,\ \rho _{\min }\approx 0.31,\ \theta \approx 29^{o} \\
\pi ^{+}\pi ^{-} &:&\ \Delta \approx -21^{o},\ \eta \approx 0.48,\ \rho
_{\min }\approx 0.35,\ \theta \approx 31^{o}
\end{eqnarray*}%
Hence we get the following bounds%
\begin{eqnarray*}
\pi ^{+}D^{-}\left( \pi ^{-}D^{+}\right) &:&\ \ \ \ \ \ \ \ \ \ \ \ 0\leq
\delta _{f,\ \bar{f}}-\Delta \leq 26^{o} \\
K^{+}\pi ^{-}\ or\ K^{0}\pi ^{+} &:&\ \ \ \ \ \ \ \ \ \ \ \ 0\leq \delta
_{f}-\Delta \leq 29^{o} \\
\pi ^{+}\pi ^{-} &:&\ \ \ \ \ \ \ \ \ \ \ \ 0\leq \delta _{f}-\Delta \leq
31^{o}
\end{eqnarray*}%
For the tree amplitude, factorization implies $\delta _{f}^{T}=0.$ We can
therefore take the point of view, the effective final state phase shift is
given by $\delta _{f}-\Delta .\ $We take the lower bounds for the tree
amplitude so that final state effective phase shift $\delta _{f}^{T}=0.$ For
the penguin we assume that the effective value of the final state phase
shift $\delta _{f}^{P}$ is near the upper bound. Thus for $\pi
^{+}D^{-}\left( \pi ^{-}D^{+}\right) ,$ $\delta _{f}^{T}=\delta _{\bar{f}%
}^{\prime T}\approx 0$; for $K^{+}\pi ^{-},$ the phase shift $\delta
_{+-}=\delta _{+-}^{P}\sim 29^{o}$ where as for $\pi ^{+}\pi ^{-},$ the
phase shift $\delta _{+-}=\delta _{+-}^{P}\sim 31^{o}.$ These phase shifts
are relavent for the Direct CP-asymmetries for $B^{0}\rightarrow K^{+}\pi
^{-}$ and $B^{0}\rightarrow \pi ^{+}\pi ^{-}$ decays.

The decay $B^{0}\rightarrow K^{+}\pi ^{-}$ is described by two amplitudes
(For details see ref.[13])%
\begin{eqnarray*}
A\left( B^{0}\rightarrow K^{+}\pi ^{-}\right) &=&-\left[ P+e^{i\gamma _{T}}%
\right] =\left\vert P\right\vert \left[ 1-re^{i\left( \gamma +\delta
_{+-}\right) }\right] \\
P &=&-\left\vert P\right\vert e^{-i\delta _{p}},\ \ T=\left\vert T\right\vert
\\
\delta _{+-} &=&\delta _{P},\text{ \ }r=\frac{\left\vert T\right\vert }{%
\left\vert P\right\vert } \\
A_{CP}\left( B^{0}\rightarrow \pi ^{-}K^{+}\right) &=&\frac{-2r\sin \gamma
\sin \delta _{+-}}{R} \\
R &=&1-2r\cos \gamma \cos \delta _{+-}+r^{2}
\end{eqnarray*}%
Neglecting the terms of order $r^{2},$%
\begin{equation*}
\tan \gamma \tan \delta _{+-}=\frac{-A_{CP}\left( B^{0}\rightarrow \pi
^{-}K^{+}\right) }{1-R}
\end{equation*}%
From the experimental values of $A_{CP}=\left( -0.097\pm 0.012\right) $ and $%
R=0.899\pm 0.048,$ with $\delta _{+-}\approx 29^{o},$ we get $\gamma =\left(
60\pm 3\right) ^{o}.$However for $\delta _{+-}\approx 20^{o},$ (which
corresponds to $\delta _{f}-\Delta \approx 20^{o};$ the value one gets for $%
\rho ^{2}=0.65$), we get $\gamma =\left( 69\pm 3\right) ^{o}.$

The phase shift $\delta _{+-}\approx \left( 20\sim 29\right) ^{o}$ for the $%
K^{+}\pi ^{-}$ is compatible with the experimental value of the direct
CP-asymmetry for $B^{0}\rightarrow K^{+}\pi ^{-}$ decay mode. For $\pi
^{+}\pi ^{-},$ $\delta _{+-}\sim 31^{o}$ is compatible with the value $%
\left( 33\pm 7%
\begin{array}{c}
+8 \\ 
-10%
\end{array}%
\right) ^{o}$ obtained by the authers of ref. [13]. In any case, our
analysis shows that the upper limit for final state stronge phase $\delta
_{f}$ is around $30^{o}$. Finally, we note that the actual value of the
effective final state phase shift $\left( \delta _{f}-\Delta \right) $
depends on one free parameter $\rho ;$ the factorization implies $\delta
_{f}^{T}=0$ $i.e.\ \left( \delta _{f}-\Delta \right) =0$ for the tree
amplitude; for the penguin amplitude, $\delta _{f}^{P}$ depends on $\rho ;$
in any case it can not be greator than the upper bound.

\section{CP Asymmetries and Strong Phases}

\textbf{Case II:}

Now, we discuss the experimental tests to verify the equality (implied by
C-invariance of S-marix) of phase shifts $\delta _{f}$ and $\delta _{\bar{f}%
} $ for the decays $B\rightarrow \pi D,\pi D^{*},\rho D$ and $%
B_{s}\rightarrow KD_{s},KD_{s}^{*},K^{*}D_{s}$.\newline

From Eqs.$\eqref{4.35b}$, we note that CP-asymetries: 
\begin{eqnarray*}
-\frac{S_{-}+S_{+}}{2} &=&\frac{2r_{D}}{1+r_{D}^{2}}\sin (2\beta +\gamma
)\cos (\delta _{f}-\delta _{\overline{f}}^{\prime }) \\
-\frac{S_{+}-S_{-}}{2} &=&\frac{2r_{D}}{1+r_{D}^{2}}\cos (2\beta +\gamma
)\sin (\delta _{f}-\delta _{\overline{f}}^{\prime })
\end{eqnarray*}%
involve dthe weak phase $2\beta +\gamma $ and strong phase $\delta
_{f}-\delta _{\overline{f}}^{\prime }.$ These asymmetries are of interst
because for 
\begin{equation*}
\delta _{f}=\delta _{\overline{f}}^{\prime },\frac{S_{+}-S_{-}}{2}=0
\end{equation*}%
and 
\begin{equation*}
-\frac{S_{-}+S_{+}}{2}=\frac{2r_{D}}{1+r_{D}^{2}}\sin (2\beta +\gamma )
\end{equation*}

Hence we can verify the equality of phases $\delta _{f}$ and $\delta _{%
\overline{f}}^{\prime }$ and determine the weak phase $2\beta +\gamma .$

For $B_{s}^{0},$ replace $r_{D}\rightarrow r_{s}$, $\delta _{f}\rightarrow
\delta _{f_{s}}$, $\delta _{\overline{f}}^{\prime }=\delta _{\overline{f}%
_{s}}^{\prime }$ and $\beta $ by $\beta _{s}.$ In standared model $\beta
_{s}=0.$

The experimental results for the B decays are as follows discussed in
section 4 
\begin{equation*}
\begin{array}{cccc}
& D^{-}\pi ^{+}\quad & D^{*-}\pi ^{+}\quad & D^{-}\rho ^{+} \\ 
\frac{S_{-}+S_{+}}{2} & -0.046\pm 0.023 & -0.037\pm 0.012 & -0.024\pm
0.031\pm 0.009 \\ 
\frac{S_{-}-S_{+}}{2} & -0.022\pm 0.021 & -0.006\pm 0.016 & -0.098\pm
0.055\pm 0.018%
\end{array}%
\end{equation*}
To determine the parameter $r_{D}$ or $r_{s}$, we assume factorization for
the tree amplitude. Factorization gives for the decays $\bar{B}%
^{0}\rightarrow D^{+}\pi ^{-},D^{*+}\pi ^{-},D^{+}\rho ^{-},D^{+}a_{1}^{-}$: 
\begin{align}
|\bar{F}_{\bar{f}}|=|\bar{T}_{\bar{f}}|& =G[f_{\pi
}(m_{B}^{2}-m_{D}^{2})f_{0}^{B-D}(m_{\pi }^{2}),2f_{\pi }m_{B}|\vec{p}%
|A_{0}^{B-D^{*}}(m_{\pi }^{2}),  \notag \\
& 2f_{\rho }m_{B}|\vec{p}|f_{+}^{B-D}(m_{\rho }^{2}),2f_{a_{1}}m_{B}|\vec{p}%
|f_{+}^{B-D}(a_{1}^{2})]  \label{c1} \\
|\bar{F}_{\bar{f}}^{^{\prime }}|=|\bar{T}_{\bar{f}}^{^{\prime }}|&
=G^{^{\prime }}[f_{D}(m_{B}^{2}-m_{\pi }^{2})f_{0}^{B-\pi
}(m_{D}^{2}),2f_{D^{*}}m_{B}|\vec{p}|f_{+}^{B-\pi }(m_{D^{*}}^{2}),  \notag
\\
& 2f_{D}m_{B}|\vec{p}|A_{0}^{B-\rho }(m_{D}^{2}),2f_{D}m_{B}|\vec{p}%
|A_{0}^{B-a_{1}}(m_{B}^{2})]  \label{c2} \\
G& =\frac{G_{F}}{\sqrt{2}}|V_{ud}||V_{cb}|a_{1},\quad G^{^{\prime }}=\frac{%
G_{F}}{\sqrt{2}}|V_{cd}||V_{ub}|  \label{c3}
\end{align}
\begin{align}
\Gamma (\bar{B}^{0}\rightarrow D^{+}\pi ^{-})&
=|V_{cb}|^{2}|f_{0}^{B-D}(m_{\pi }^{2})|^{2}(2.281\times 10^{-9})MeV  \notag
\\
\Gamma (\bar{B}^{0}\rightarrow D^{*+}\pi ^{-})&
=|V_{cb}|^{2}|A_{0}^{B-D^{*}}(m_{\pi }^{2})|^{2}(2.129\times 10^{-9})MeV 
\notag \\
\Gamma (\bar{B}^{0}\rightarrow D^{+}\rho ^{-})&
=|V_{cb}|^{2}|f_{+}^{B-D}(m_{\rho }^{2})|^{2}(5.276\times 10^{-9})MeV  \notag
\\
\Gamma (\bar{B}^{0}\rightarrow D^{+}a_{1}^{-})&
=|V_{cb}|^{2}|f_{+}^{B-D}(m_{a_{1}}^{2})|^{2}(5.414\times 10^{-9})MeV
\label{c4}
\end{align}

\begin{table}[ht]
\centering  
\begin{tabular}{|c|c|c|c|}
\hline
Decay & Decay Width $(10^{-9}$ MeV $\times |V_{cb}|^{2}$) & Form Factor & 
Form Factors $h(w^{(\ast)})$ \\ \hline
$\bar{B}^{0} \rightarrow D^{+} \pi^{-}$ & $(2.281) |f_{0}^{B-D}
(m_{\pi}^{2})|^{2}$ & $0.58 \pm 0.05$ & $0.51 \pm 0.03$ \\ \hline
$\bar{B}^{0} \rightarrow D^{\ast +} \pi^{-}$ & $(2.129) |A_{0}^{B-D \ast}
(m_{\pi}^{2})|^{2}$ & $0.61 \pm 0.04$ & $0.54 \pm 0.03$ \\ \hline
$\bar{B}^{0} \rightarrow D^{+} \rho^{+}$ & $(5.276) |f_{+}^{B-D}
(m_{\rho}^{2})|^{2}$ & $0.65 \pm 0.11 $ & $0.57 \pm 0.10$ \\ \hline
$\bar{B}^{0} \rightarrow D^{+} a_{1}$ & $(5.414) |f_{+}^{B-D}
(m_{a_{1}}^{2})|^{2}$ & $0.57 \pm 0.31 $ & $0.50 \pm 0.27$ \\ \hline
\end{tabular}%
\caption{Form Factors}
\label{tab:1}
\end{table}
The decay widths for the above channels are given in the table 1

where we have used 
\begin{equation*}
a_{1}^{2}|V_{ud}|^{2}\approx 1,\quad f_{\pi }=131MeV,\quad f_{\rho
}=209MeV,\quad f_{a_{1}}=229MeV
\end{equation*}

Using the experimental branching ratios and 
\begin{equation}
|V_{cb}|=(38.3\pm 1.3)\times 10^{-3}  \label{c5}
\end{equation}

we obtain the corresponding form factors given in Table 1. 
\begin{align}
|f_{0}^{B-D}(m_{\pi}^{2})| &= 0.58 \pm 0.05  \notag \\
|A_{0}^{B-D^{\ast}}(m_{\pi}^{2})| &= 0.61 \pm 0.04  \notag \\
|f_{+}^{B-D}(m_{\rho}^{2})| &= 0.65 \pm 0.11  \notag \\
|f_{+}^{B-D}(m_{a_{1}}^{2})| &= 0.57 \pm 0.31  \label{c6}
\end{align}

In terms of Isgur Wise variables: 
\begin{equation}
\omega =v\cdot v^{^{\prime }},\quad v^{2}=v^{^{\prime }2}=1,\quad
t=q^{2}=m_{B}^{2}+m_{D^{*}}^{2}-2m_{B}m_{D^{*}}\omega  \label{c7}
\end{equation}

the form factors can be put in the following form 
\begin{align}
f_{+}^{B-D}(t) &= \frac{m_{B} + m_{D}}{2\sqrt{m_{B}m_{D}}} h_{+}(\omega),
\quad f_{0}^{B-D}(t) = \frac{\sqrt{m_{B}m_{D}}}{m_{B} + m_{D}} (1 +
\omega)h_{0}(\omega)  \notag \\
A_{2}^{B-D^{\ast}} (t) &= \frac{m_{B} + m_{D^{\ast}}}{2\sqrt{%
m_{B}m_{D^{\ast}}}} (1 + \omega) h_{A_{2}}(\omega), \quad A_{0}^{B-D^{\ast}}
(t) = \frac{m_{B} + m_{D^{\ast}}}{2\sqrt{m_{B}m_{D^{\ast}}}} h_{A_{0}}
(\omega)  \notag \\
A_{1}^{B-D^{\ast}} (t) &= \frac{\sqrt{m_{B}m_{D^{\ast}}}}{m_{B} +
m_{D^{\ast}}} (1 + \omega) h_{A_{1}} (\omega)  \label{c8}
\end{align}

Heavy Quark Effective Theory (HQET) gives: 
\begin{equation*}
h_{+} (\omega) = h_{0} (\omega) = h_{A_{0}} (\omega) = h_{A_{1}} (\omega) =
h_{A_{2}} (\omega) = \zeta(\omega)
\end{equation*}

where $\zeta(\omega)$ is Isgur-Wise form factor, with normalization $%
\zeta(1) = 1$. For 
\begin{align*}
t &= m_{\pi}^{2}, m_{\rho}^{2}, m_{a_{1}}^{2} \\
\omega^{\ast} &= 1.589(1.504), 1.559, 1.508
\end{align*}

we get the form factors h's given in Table 1.

In reference , the value quoted for $h_{A_{1}}(\omega _{max}^{*})$ is 
\begin{equation}
|h_{A_{1}}(\omega _{max}^{*})|=0.52\pm 0.03  \label{c10}
\end{equation}

Since $\omega_{max}^{\ast} = 1.504$, the value for $|h_{A_{0}}(max)|$
obtained in Table 1 is in remarkable agreement with the value given in Eq. $%
\eqref{c10}$that assumption for $B^{0} \rightarrow \pi D^{(\ast)}$ decays is
experimentally on solid footing and is in agreement with HQET.

From Eqs. $\eqref{c1}$ and $\eqref{c2}$, we obtain 
\begin{align}
r_{D} &= \lambda^{2} R_{b} \frac{|\bar{T}_{f}^{^{\prime}}|}{|\bar{T}_{\bar{f}%
}|}  \notag \\
&= \lambda^{2} R_{b} \left[ \frac{f_{D}(m_{B}^{2} - m_{\pi}^{2})
f_{0}^{B-\pi} (m_{D}^{2})}{f_{\pi}(m_{B}^{2} - m_{D}^{2}) f_{0}^{B-D}
(m_{\pi}^{2})}, \quad \frac{f_{D^{\ast}} f_{+}^{B-\pi}(m_{D^{\ast}}^{2})} {%
f_{\pi} A_{0}^{B-D}(m_{\pi}^{2})}, \quad \frac{f_{D}
A_{0}^{B-\rho}(m_{D}^{2}) } {f_{\rho} f_{+}^{B-D}(m_{\rho^{2}})} \right]
\label{c11}
\end{align}

where 
\begin{equation}
\frac{|V_{ub}| |V_{cd}|}{|V_{cb}| |V_{ud}|} = \lambda^{2} R_{b} \approx
(0.227)^{2} (0.40) \approx 0.021  \label{c12}
\end{equation}

To determine $r_{D}$, we need information for the form factors $f_{0}^{B-\pi
}(m_{D}^{2}),f_{+}^{B-\pi }(m_{D}^{2}),A_{0}^{B-\rho }(m_{D}^{2})$. For
these form factors, we use the following values: 
\begin{align*}
A_{0}^{B-\rho }(0)& =0.30\pm 0.03,A_{0}^{B-\rho }(m_{D}^{2})=0.38\pm 0.04 \\
f_{+}^{B-\pi }(0)& =f_{0}^{B-\pi }(0)=0.26\pm 0.04,\quad f_{+}^{B-\pi
}(m_{D^{\ast }}^{2})=0.32\pm 0.05,\quad f_{0}^{B-D}(m_{D}^{2})=0.28\pm 0.04
\end{align*}

Along with the remaining form factors given in Table, we obtain 
\begin{equation}
r_{D}=[0.018\pm 0.002,\quad 0.017\pm 0.003,\quad 0.012\pm 0.002]  \label{c13}
\end{equation}

The above value for $r_{D}^{\ast}$ gives 
\begin{equation}
- \left( \frac{S_{+} + S_{-}}{2} \right)_{D^{\ast} \pi} = 2 (0.017 \pm
0.003) \sin (2\beta + \gamma)  \label{c15}
\end{equation}

The experimental value of the CP asymmetry for $B^{0}\rightarrow D^{*}\pi $
decay has the least error. Hence we obtain the following bounds 
\begin{align}
\sin (2\beta +\gamma )& >0.69  \label{c16} \\
44^{\circ }& \leq (2\beta +\gamma )\leq 90^{\circ } \\
\text{or}\quad 90^{\circ }& \leq (2\beta +\gamma )\leq 136^{\circ }
\end{align}

Selecting the second solution, and using $\beta \approx 43^{\circ }$, we get 
\begin{equation}
\gamma =(70\pm 23)^{\circ }  \label{c19}
\end{equation}

To end this section, we discuss the decays $\bar{B}_{s}^{0}\rightarrow
D_{s}^{+}K^{-},D_{s}^{*+}K^{-}$ for which no experimental data is available.
However, using facorization, we get 
\begin{align}
\Gamma (\bar{B}_{s}^{0}\rightarrow D_{s}^{+}K^{-})& =(1.75\times
10^{-10})|V_{cb}f_{0}^{B_{s}-D_{s}}(m_{K}^{2})|^{2}MeV  \label{c31} \\
\Gamma (\bar{B}_{s}^{0}\rightarrow D_{s}^{*+}K^{-})& =(1.57\times
10^{-10})|V_{cb}A_{0}^{B_{s}-D_{s}^{*}}(m_{K}^{2})|^{2}MeV  \label{c32}
\end{align}

SU(3) gives 
\begin{align}
|V_{cb} f_{0}^{B_{s} - D_{s}} (m_{K}^{2})|^{2} &\approx |V_{cb}| |f_{0}^{B -
D} (m_{\pi}^{2})|^{2} = (0.50 \pm 0.04) \times 10^{-3}  \notag \\
|V_{cb} A_{0}^{B_{s} - D_{s}^{\ast}} (m_{K}^{2})|^{2} &\approx |V_{cb}|
|A_{0}^{B - D} (m_{\pi}^{2})|^{2} = (0.56 \pm 0.04) \times 10^{-3}
\label{c33}
\end{align}

From the above equations, we get the following branching ratios 
\begin{equation}
\frac{\Gamma (\bar{B_{s}}^{0} \rightarrow D_{s}^{(\ast) + }K^{-})} {\Gamma_{%
\bar{B}_{s}^{0}}} = (1.94 \pm 0.07) \times 10^{-4} [(1.96 \pm 0.07) \times
10^{-4}]  \label{c34}
\end{equation}

For $\bar{B}_{s}^{0} \rightarrow D_{s}^{\ast +} K^{-}$ 
\begin{equation}
r_{s} = R_{b} \left[ \frac{f_{D_{s}^{\ast}} f_{+}^{B_{s} - K}
(m_{D_{s}^{\ast}}^{2})}{f_{K} A_{0}^{B_{s} - D_{s}^{\ast}} (m_{K}^{2})} %
\right]  \label{c35}
\end{equation}

Hence we get 
\begin{align}
-(\frac{S_{+} + S_{-}}{2})_{D_{s}^{\ast} K} &= (0.41 \pm 0.08) \sin
(2\beta_{s} + \gamma)  \notag \\
&= (0.41 \pm 0.08) \sin \gamma  \label{37}
\end{align}

where we have used 
\begin{align}
R_{b}& =0.40,\quad \frac{f_{D_{s}}}{f_{K}}=\frac{f_{D_{s}^{*}}}{f_{K}}%
=1.75\pm 0.06,\quad f_{+}^{B_{s}-K}(m_{D_{s}^{*}}^{2})=0.34\pm 0.06  \notag
\\
A_{0}^{B_{s}-D_{s}^{*}}(m_{K}^{2})& =A_{0}^{B_{s}-D_{s}^{*}}(0)=\frac{%
m_{B_{s}}+m_{D_{s}^{*}}}{2\sqrt{m_{B_{s}m_{D_{s}^{*}}}}}\left[ h_{0}(\omega
_{s}^{*}=1.453)=0.52\pm .03\right]  \notag \\
& =0.58\pm 0.03  \label{c36}
\end{align}

\textbf{Case III}

We now confine ourselves to $B^{0}(\bar{B}^{0})\rightarrow \rho ^{-}\pi
^{+},\rho ^{+}\pi ^{-}(\rho ^{+}\pi ^{-},\rho ^{-},\pi ^{+})$ decays only
[13,14]. The experimental results for these decays are [6] as 
\begin{align}
\Gamma & =R_{f}+R_{\bar{f}}=(22.8\pm 2.5)\times 10^{-6}  \label{ccc19} \\
A_{CP}^{f}& =-0.16\pm 0.23,\quad A_{CP}^{\bar{f}}=0.08\pm 0.12  \label{ccc20}
\\
C& =0.01\pm 0.14,\quad \Delta C=0.37\pm 0.08  \label{ccc21} \\
S& =0.01\pm 0.09,\quad \Delta S=-0.05\pm 0.10  \label{ccc22}
\end{align}

With the above values, it is hard to draw any reliable conclusion.
Neglecting the term $A_{CP}C$ in Eqs. $\eqref{ccc9}$ and $\eqref{ccc10}$, we
get 
\begin{align}
R_{\bar{f},f}& =\frac{1}{2}\Gamma (1\pm \Delta C)  \label{ccc23} \\
R_{\bar{f}}-R_{f}& =\Delta C
\end{align}

Using the above value for $\Delta C$, we obtain 
\begin{align}
R_{\bar{f}} &= (15.6 \pm 1.7) \times 10^{-6}  \notag \\
R_{f} &= (7.2 \pm 0.8) \times 10^{-6}  \label{ccc25}
\end{align}

We analyze these decays by assuming factorization for the tree graphs$\left[ 
\text{19}\right] $. This assumption gives 
\begin{align}
T_{\bar{f}}& =\bar{T}_{f}\sim 2m_{B}f_{\rho }|\vec{p}|f_{+}(m_{\rho }^{2})
\label{ccc26} \\
T_{f}& =\bar{T}_{\bar{f}}\sim 2m_{B}f_{\pi }|\vec{p}|A_{0}(m_{\pi }^{2})
\label{ccc27}
\end{align}

Using $f_{+}(m_{\rho }^{2})\approx 0.26\pm 0.04$ and $A_{0}(m_{\pi
}^{2})\approx A_{0}(0)=0.29\pm 0.03$ and $|V_{ub}|=(3.5\pm 0.6)\times
10^{-3} $, we get the following values for the tree amplitude contribution
to the branching ratios 
\begin{align}
\Gamma _{\bar{f}}^{\text{tree}}& =(15.6\pm 1.1)\times 10^{-6}\equiv |T_{\bar{%
f}}|^{2}  \label{ccc28} \\
\Gamma _{f}^{\text{tree}}& =(7.6\pm 1.4)\times 10^{-6}\equiv |T_{f}|^{2}
\label{ccc29} \\
t& =\frac{T_{f}}{T_{\bar{f}}}=\frac{f_{\pi }A_{0}(m_{\pi }^{2})}{f_{\rho
}f_{+}(m_{\rho }^{2})}=0.70\pm 0.12
\end{align}

Now 
\begin{align}
B_{\bar{f}} &= \frac{R_{\bar{f}}}{|T_{\bar{f}}|^{2}} = 1 - 2 r_{\bar{f}}
\cos \alpha \cos \delta_{\bar{f}} + r_{\bar{f}}^{2}  \label{ccc31} \\
B_{f} &= \frac{R_{f}}{|T_{f}|^{2}} = 1 - 2 r_{f} \cos \alpha \cos \delta_{f}
+ r_{f}^{2}  \label{ccc32}
\end{align}

Hence from Eqs. $\eqref{ccc25}$ and $\eqref{ccc29}$, we get 
\begin{align}
B_{\bar{f}} &= 1.00 \pm 0.12  \notag \\
B_{f} &= 0.95 \pm 0.11  \label{ccc33}
\end{align}

In order to take into account the contribution of penguin diagram, we
introduce the angles $\alpha _{eff}^{f,\bar{f}}$ , defined as follows 
\begin{align}
e^{i\beta }A_{f,\bar{f}}& =|A_{f,\bar{f}}|e^{-i\alpha _{eff}^{f,\bar{f}}} 
\notag \\
e^{-i\beta }\bar{A}_{f,\bar{f}}& =|\bar{A}_{f,\bar{f}}|e^{i\alpha _{eff}^{f,%
\bar{f}}}  \label{ccc34}
\end{align}

With this definition, we separate out tree and penguin contributions: 
\begin{align}
e^{i \beta} A_{f, \bar{f}} - e^{-i \beta} \bar{A}_{f, \bar{f}} &= |A_{f, 
\bar{f}}| e^{-i \alpha^{f, \bar{f}}} - |\bar{A}_{f, \bar{f}}| e^{i\alpha^{f, 
\bar{f}}}  \notag \\
&= 2 i T_{f, \bar{f}} \sin \alpha  \label{ccc35} \\
e^{i (\alpha + \beta)} A_{f, \bar{f}} - e^{- i (\alpha + \beta)} \bar{A}_{f, 
\bar{f}} &= |A_{f, \bar{f}}| e^{- i (\alpha_{eff}^{f, \bar{f}} - \alpha)} 
\notag \\
&= (2 i T_{f, \bar{f}} \sin \alpha) r_{f, \bar{f}} e^{i \delta_{f, \bar{f}}}
\notag \\
&= 2 i P_{f, \bar{f}} \sin \alpha  \label{ccc36}
\end{align}

From Eq. $\eqref{ccc35}$, we get 
\begin{align}
2\frac{|T_{f,\bar{f}}|^{2}}{R_{f,\bar{f}}}\sin ^{2}\alpha & \equiv \frac{%
2\sin ^{2}\alpha }{B_{f,\bar{f}}}=1-\sqrt{1-A_{CP}^{f,\bar{f}2}}\cos 2\alpha
_{eff}^{f,\bar{f}}  \label{ccc37} \\
\sin 2\delta _{f,\bar{f}}^{T}& =-A_{CP}^{f,\bar{f}}\frac{\sin 2\alpha
_{eff}^{f,\bar{f}}}{1-\sqrt{1-A_{CP}^{f,\bar{f}2}}\cos 2\alpha _{eff}^{f,%
\bar{f}}}  \label{ccc38a} \\
\cos 2\delta _{f,\bar{f}}^{T}& =\frac{\sqrt{1-A_{CP}^{f,\bar{f}2}}-\cos
2\alpha _{eff}^{f,\bar{f}}}{1-\sqrt{1-A_{CP}^{f,\bar{f}2}}\cos 2\alpha
_{eff}^{f,\bar{f}}}  \label{ccc38b} \\
&
\end{align}

From Eqs. $\eqref{ccc35}$ and $\eqref{ccc36}$, we get 
\begin{align}
r_{f,\bar{f}}^{2}& =\frac{1-\sqrt{1-A_{CP}^{f,\bar{f}2}}\cos (2\alpha
_{eff}^{f,\bar{f}}-2\alpha )}{1-\sqrt{1-A_{CP}^{f,\bar{f}2}}\cos 2\alpha
_{eff}^{f,\bar{f}}}  \label{ccc39} \\
r_{f,\bar{f}}\cos \delta _{f,\bar{f}}& =\frac{\cos \alpha -\sqrt{1-A_{CP}^{f,%
\bar{f}2}}\cos (2\alpha _{eff}^{f,\bar{f}}-\alpha )}{1-\sqrt{1-A_{CP}^{f,%
\bar{f}2}}\cos 2\alpha _{eff}^{f,\bar{f}}}  \label{ccc40} \\
r_{f,\bar{f}}\sin \delta _{f,\bar{f}}& =\frac{-A_{CP}^{f,\bar{f}}/\sin
\alpha }{1-\sqrt{1-A_{CP}^{f,\bar{f}2}}\cos 2\alpha _{eff}^{f,\bar{f}}}
\label{ccc41}
\end{align}

Now factorization implies [23] 
\begin{equation}
\delta _{f}^{T}=0=\delta _{\bar{f}}^{T}  \label{ccc42}
\end{equation}

Thus in the limit $\delta_{f}^{T} \rightarrow 0$, we get for Eq. $%
\eqref{ccc38b}$ 
\begin{align}
\cos 2 \alpha_{eff}^{f, \bar{f}} &= -1, \qquad \alpha_{eff}^{f, \bar{f}} =
90^{\circ}  \label{ccc43} \\
r_{f, \bar{f}} \cos \delta_{f, \bar{f}} &= \cos \alpha  \label{ccc44} \\
r_{f, \bar{f}} \sin \delta_{f, \bar{f}} &= \frac{-A_{CP}^{f, \bar{f}} / \sin
\alpha}{1 + \sqrt{1 - A_{CP}^{f, \bar{f} 2}}}  \label{ccc45} \\
r_{f, \bar{f}}^{2} &= \frac{1 + \sqrt{1 - A_{CP}^{f, \bar{f} 2}} \cos 2
\alpha}{1 + \sqrt{1 - A_{CP}^{f, \bar{f} 2}}}  \label{ccc47} \\
&\approx \cos^{2} \alpha + \frac{1}{4} A_{CP}^{f, \bar{f} 2} \sin^{2} \alpha
\label{ccc48}
\end{align}


The solution of Eq. $\eqref{ccc44}$ is graphically shown in Fig. 7 for $%
\alpha$ in the range $80^{\circ} \leq \alpha < 103^{\circ}$ for $r_{f, \bar{f%
}} = 0.10, 015, 0.20, 0.25, 0.30$. From the figure, the final state phases $%
\delta_{f, \bar{f}}$ for various values of $r_{f, \bar{f}}$ can be read for
each value of $\alpha$ in the above range. Few examples are given in Table 2

\begin{table}[ht]
\centering  
\begin{tabular}{|c|c|c|c|}
\hline
$\alpha$ & $r_{f}$ & $\delta_{f}$ & $A_{CP}^{f} \approx -2 r_{f} \sin
\delta_{f} \sin \alpha$ \\ \hline
$80^{\circ}$ & 0.20 & $29^{\circ}$ & -0.19 \\ \hline
& 0.25 & $46^{\circ}$ & -0.36 \\ \hline
$82^{\circ}$ & 0.15 & $22^{\circ}$ & -0.11 \\ \hline
& 0.20 & $46^{\circ}$ & -0.28 \\ \hline
$85^{\circ}$ & 0.10 & $29^{\circ}$ & -0.10 \\ \hline
& 0.15 & $54^{\circ}$ & -0.24 \\ \hline
$86^{\circ}$ & 0.10 & $46^{\circ}$ & -0.14 \\ \hline
& 0.15 & $62^{\circ}$ & -0.26 \\ \hline
$88^{\circ}$ & 0.10 & $70^{\circ}$ & -0.19 \\ \hline
\end{tabular}%
\caption{}
\label{tab:2}
\end{table}

For $\alpha >90^{\circ }$, change $\alpha \rightarrow \pi -\alpha $, $\delta
_{f}\rightarrow \pi -\delta _{f}$. For example, for $\alpha =103^{\circ }$ 
\begin{align*}
r_{f}& =0.25,\quad \delta _{f}=154^{\circ },\quad A_{CP}^{f}\approx -0.22 \\
r_{f}& =0.30,\quad \delta _{f}=138^{\circ },\quad A_{CP}^{f}\approx -0.40
\end{align*}

These examples have been selected keeping in view that final state phases $%
\delta_{f, \bar{f}}$ are not too large. For $A^{f, \bar{f}}_{CP}$, we have
used Eq. $\eqref{ccc45}$ neglecting the second order term. An attractive
option is $A_{CP}^{f} = A_{CP}^{\bar{f}}$ for each value of $\alpha$;
although $A_{CP}^{f} \neq A_{CP}^{\bar{f}}$ is also a possibility. $%
A^{f}_{CP} = A_{CP}^{\bar{f}}$ implies $r_{f} = r_{\bar{f}}, \delta_{f} =
\delta_{\bar{f}}$.

Neglecting terms of order $r_{f, \bar{f}}^{2}$, we have 
\begin{align}
A_{CP} \approx \frac{2 \sin \alpha (r_{\bar{f}} \sin \delta_{\bar{f}} -
t^{2} r_{f} \sin \delta_{f} )} {1 + t^{2}} = - \frac{A_{CP}^{\bar{f}} -
t^{2} A_{CP}^{f}}{1 + t^{2}}  \label{cccc1} \\
C \approx - \frac{2 t^{2}}{(1 + t)^{2}} (A_{CP}^{\bar{f}} + A_{CP}^{f})
\label{cccc2} \\
\Delta C \approx \frac{1 - t^{2}}{1 + t^{2}} - \frac{4 t^{2} \cos \alpha}{(1
+ t^{2})^{2}} (r_{\bar{f}} \cos \delta_{\bar{f}} - r_{f} \cos \delta_{f})
\label{cccc3}
\end{align}

Now the second term in Eq. $\eqref{cccc3}$ vanishes and using the value of $%
t $ given in Eq. $\eqref{ccc30}$, we get 
\begin{equation}
\Delta C \approx 0.34 \pm 0.06  \label{cccc4}
\end{equation}

Assuming $A_{CP}^{\bar{f}} = A_{CP}^{f}$, we obtain 
\begin{align}
A_{CP} &= - \frac{1 - t^{2}}{1 + t^{2}} A_{CP}^{\bar{f}}  \notag \\
&= (0.34 \pm 0.06) (-A_{CP}^{\bar{f}})  \label{cccc5} \\
C &\approx - \frac{4 t^{2}}{(1 + t^{2})^{2}} A_{CP}^{\bar{f}} \approx -
(0.88 \pm 0.14) A_{CP}^{\bar{f}}  \label{cccc6}
\end{align}

Finally the CP asymmetries in the limit $\delta _{f,\bar{f}}^{T}\rightarrow
0 $ 
\begin{align}
S_{\bar{f}}=S+\Delta S& =\frac{2\text{Im}[e^{2i\phi _{M}}A_{\bar{f}^{*}}\bar{%
A}_{\bar{f}}]}{\Gamma (1+A_{CP})}  \notag \\
& =\sqrt{1-C_{\bar{f}}^{2}}\sin (2\alpha _{eff}^{\bar{f}}+\delta )  \notag \\
& =-\sqrt{1-C_{\bar{f}}^{2}}\cos \delta  \label{cccc7} \\
S_{f}=S-\Delta S& =\frac{2\text{Im}[e^{2i\phi _{M}}A_{f}^{*}\bar{A}_{f}]}{%
\Gamma (1-A_{CP})}  \notag \\
& =\sqrt{1-C_{f}^{2}}\sin (2\alpha _{eff}^{\bar{f}}-\delta )  \notag \\
& =\sqrt{1-C_{f}^{2}}\cos \delta  \label{cccc8}
\end{align}

The phase $\delta$ is defined as 
\begin{equation}
\bar{A}_{\bar{f}} = \frac{| \bar{A}_{\bar{f} |}}{| \bar{A}_{f} |} \bar{A}%
_{f} e^{i \delta}  \label{cccc9}
\end{equation}

Hence we have 
\begin{equation*}
\frac{S+\Delta S}{S-\Delta S}=-\frac{\sqrt{1-C_{\bar{f}}^{2}}}{\sqrt{%
1-C_{f}^{2}}}
\end{equation*}


\section{Conclusion}

In weak interaction, both P and C are violated but CP is conserved by the
weak interaction Lagrangian. Hence for $X^{0}-\bar{X}^{0}$ complex $%
(X^{0}=K^{0},B^{0},B_{s}^{0})$; the mass matrix is not diagonal in $%
|X^{0}\rangle $ and $|\bar{X}^{0}\rangle $ basis. However, assuming $CP$
conservation, the $CP$ eigenstates $|X_{1}^{0}\rangle $ and $%
|X_{2}^{0}\rangle $ can be mass eigenstates and hence mass matrix is
diagonal in this basis. The two sets of states are related to each other by
superposition principle of quantum mechanics. This gives rise to quantum
mechanical interference so that even if we start with a state $|X^{0}\rangle 
$, the time evolution of this state can generate the state $|X^{0}\rangle $.
This is a source of mixing induced $CP$ violation. However, both in $K^{0}-%
\bar{K}^{0}$ and $B^{0}-\bar{B}^{0}$ complex, the mass eigenstates $%
|K_{S}^{0}\rangle $, $|K_{L}^{0}\rangle $ and $|B_{L}^{0}\rangle $, $%
|B_{H}^{0}\rangle $ are not $CP$ eigenstates. In the case of $K^{0}-\bar{K}%
^{0}$ complex, there is a small admixture of wrong $CP$ state characterized
by a small parameter $\epsilon $, which gives rise to the $CP$ violating
decay $K_{L}^{0}\rightarrow \pi ^{+}\pi ^{-}$. This was the first $CP$
violating decay observed experimentally. For $B^{0}-\bar{B}^{0}$ complex,
the mismatch between mass eigenstates and $CP$ eigenstates $%
|B_{1}^{0}\rangle $ and $|B_{2}^{0}\rangle $ is given by the phase factor $%
e^{2i\phi _{M}}$ where the phase factor is $\phi _{M}=-\beta $ in the
standard model viz. one of the phases in the CKM matrix. For $B_{s}^{0}-\bar{%
B}_{s}^{0}$, there is no mismatch between $CP$ eigenstates $%
|B_{1s}^{0}\rangle $ and $|B_{2s}^{0}\rangle $ and the mass eigenstates.
There is no extra phase available in CKM matrix, with three generations of
quarks to accomodate more than two independent phases $\beta $ and $\gamma $%
; the unitarity of CKM matrix requires $\alpha +\beta +\gamma =\pi $.

The quantum mechanical interference gives rise to non zero mass differences $%
\Delta m_{K}$, $\Delta m_{B}$ and $\Delta m_{B_{s}}$ between mass
eigenstates. The mixing induced $CP$ violation involves these mass
differences.

The $CPT$ invariance plays an important role in $CP$ violation in weak
decays. $CPT$ invariance gives 
\begin{equation*}
\bar{A}_{\bar{f}}=\eta _{f}e^{2i\delta _{f}}A_{f}^{\ast },\quad
A_{f}=e^{i\delta _{f}}e^{i\phi }|A_{f}|
\end{equation*}

where $A_{f}$ and $\bar{A}_{\bar{f}}$ are the amplitudes for the decays $%
X\rightarrow f$ and $\bar{X}\rightarrow \bar{f}$, the states $|f\rangle $
and $|\bar{f}\rangle $ being $CP$ conjugate of each other. For direct $CP$
violation, at least two amplitudes with different weak phase are required: 
\begin{equation*}
A_{f}=A_{1f}+A_{2f}
\end{equation*}

$CPT$ gives: 
\begin{align*}
\bar{A}_{\bar{f}} &= e^{2i\delta_{1f}} A_{1f}^{\ast} + e^{2i\delta_{2f}}
A_{2f}^{\ast} \\
A_{if} &= e^{i\delta_{if}} e^{i\phi_{i}} | A_{if} |
\end{align*}

where $(\delta_{1f}, \delta_{2f})$, $(\phi_{1}, \phi_{2})$ are strong final
state phases and the weak phases respectively. Thus the direct $CP$
violation is given by 
\begin{equation*}
A_{CP} = \frac{\bar{\Gamma} (\bar{X} \rightarrow \bar{f}) - \Gamma (X
\rightarrow f)}{\bar{\Gamma} (\bar{X} \rightarrow \bar{f}) + \Gamma (X
\rightarrow f)}
\end{equation*}

where $\delta_{f} = \delta_{2f} - \delta_{1f}$, $\phi = \phi_{2} - \phi_{1}$%
. Hence the necessary condition for non-zero direct $CP$ violation is $%
\delta_{f} \neq 0$ and $\phi \neq 0$.

In section 2, the $CP$ violation due to mismatch between $CP$ eigenstates $%
|K_{1}^{0}\rangle $, $|K_{2}^{0}\rangle $ and mass eigenstates $%
|K_{S}^{0}\rangle $ and $|K_{L}^{0}\rangle $ in terms of the parameter $%
\epsilon $ and direct $CP$ violation due to different weak phases bewteen
the cecay amplitudes $A_{0}$ and $A_{2}$ are discussed.

\textbf{Section 4:}

\textbf{Case I}

The $CP$ violation for $B^{0}\rightarrow f$ decay where $|\bar{f}\rangle
=CP|f\rangle =|f\rangle $ are discussed. In particular for the decay $%
B^{0}\rightarrow J/\psi K_{S}^{0}$ described by a single amplitude $A_{f}$,
the $CP$ asymmetry is given by 
\begin{equation*}
A_{J/\psi K_{S}}=-\sin 2\beta \frac{(\Delta m_{B}/\Gamma )}{1+(\Delta
m_{B}/\Gamma )}
\end{equation*}

It is a good illustration of $CP$ violation due to mismatch between mass and 
$CP$ eigenstates, involving the mixing parameter $\Delta m_{B}$. From the
experimental values of $A_{J/\psi K_{S}}$ and $(\Delta m/\Gamma)_{B^{0}}$,
the weak phase $2\beta$ is found to be $(43 \pm 3)^{\circ}$. Corresponding
to $B^{0} \rightarrow J/\psi K^{0}_{S}$, we have $B^{0}_{S} \rightarrow
J/\psi \phi$ and for this decay 
\begin{equation*}
A_{J/\psi \phi} = - \sin 2\beta_{s} \frac{(\Delta m_{B_{S}^{0}} / \Gamma_{S})%
}{1 + (\Delta m_{B_{S}^{0}}/\Gamma_{S})^{2}}
\end{equation*}

Any finite value of $A_{J/\psi \phi}$ would imply $\beta_{s} \neq 0$ in
contradiction with the standard model.

In this section for the case (\textbf{i}), both direct and mixing induced $%
CP $ violation viz. $A_{CP}$, $C_{f}$ and $S_{f}$ for $B^{0}\rightarrow \pi
^{+}\pi ^{-}$ described by two amplitudes $T$ and $P_{t}$ given by tree and
penguin diagrams is discussed. We find $C_{\pi \pi }=-A_{CP}(\pi \pi )$ and $%
S_{\pi \pi }$ is essentially given by 
\begin{equation*}
S_{\pi \pi }\approx (\sin 2\alpha +2r\cos \delta \sin \alpha \cos 2\alpha
),\ \ \ \ \ \ \ \ \ \ \ r=\frac{R_{t}}{R_{b}}\frac{|P_{t}|}{|T|}
\end{equation*}

$S_{\pi \pi }\neq 0$ even when final state phase $\delta =0$.

\textbf{Case II}

We consider the cdecays described by two independent decay amplitudes $A_{f}$
and $A_{\bar{f}}^{^{\prime }}$ with different weak phases $(O$ and $\gamma )$
where the final states $|f\rangle $ and $|\bar{f}\rangle $ are $C$ and $CP$
conjugate of each other such as the states $D^{(\ast )-}\pi ^{+}$ $(D^{(\ast
)+}\pi ^{-})$, $D_{s}^{(\ast )-}K^{+}$ $(D_{s}^{(\ast )+}K^{-})$, $D^{-}\rho
^{+}$ $(D^{+}\rho ^{-})$.

It is argued in section 5, that $C$ and $CP$ invariance of hadronic
interactions imply $\delta_{f} = \delta_{\bar{f}}^{^{\prime}}$.

As discussed in section 6, the equality of phases $\delta_{f} = \delta_{\bar{%
f}}^{^{\prime}}$ implies that time-dependent $CP$ asymmetries: 
\begin{align*}
- \left( \frac{S_{+} + S_{-}}{2} \right) &= \frac{2 r_{D^{(\ast)}}}{1 + 2
r^{2}_{D^{(\ast)}}} \sin (2\beta + \gamma) \\
\frac{S_{+} - S_{-}}{2} &= 0
\end{align*}

It is further shown that from the experimental value of $\frac{S_{+} + S_{-}%
}{2}$ for $B^{0} \rightarrow D^{\ast -} \pi^{+}$ 
\begin{align*}
\sin (2\beta +\gamma )& >0.69 \\
44^{\circ }\leq 2\beta +\gamma \leq 90^{\circ }\quad & or\quad 90^{\circ
}\leq 2\beta +\gamma \leq 136^{\circ }
\end{align*}

Selecting the second solution and using $2\beta \approx 43^{\circ}$, we get 
\begin{equation*}
\gamma = (70 \pm 23)^{\circ}
\end{equation*}

Using $SU(3)$, for the form factors for $B_{s}^{0}\rightarrow D^{\ast
-}K^{+} $, we predict 
\begin{equation*}
-\left( \frac{S_{+}+S_{-}}{2}\right) =(0.41\pm 0.08)\sin (2\beta _{s}+\gamma
)
\end{equation*}

In the standard model $\beta _{s}=0$.

\textbf{Case III}

For the case (\textbf{III}) for which $A_{f}\neq A_{\bar{f}}$ such as $%
B^{0}\rightarrow \rho ^{+}\pi ^{-}:A_{\bar{f}}$ and $B^{0}\rightarrow \rho
^{-}\pi ^{+}:A_{f}$ where $A_{f,\bar{f}}$ are given by tree amplitude $%
e^{i\gamma }T_{f,\bar{f}}$ and penguin amplitude $e^{-i\beta }P_{f,\bar{f}}$
are discussed.

In section 6 case (\textbf{iii}), the factorization for the tree graph
implies $\delta _{f}^{T}\approx \delta _{\bar{f}}^{T}\approx 0$. In the
limit $\delta _{f,\bar{f}}^{T}\rightarrow 0$, it is shown that 
\begin{align*}
r_{f,\bar{f}}\cos \delta _{f,\bar{f}}& =\cos \alpha \\
r_{f,\bar{f}}^{2}& \approx \cos ^{2}\alpha +A_{CP}^{f,\bar{f}2}\sin
^{2}\alpha
\end{align*}

Finally, in the limit $\delta _{f,\bar{f}}^{T}\rightarrow 0$, we get 
\begin{equation*}
\frac{S_{\bar{f}}}{S_{f}}=\frac{S+\Delta S}{S-\Delta S}=-\sqrt{\frac{1-C_{%
\bar{f}}^{2}}{1-C_{f}^{2}}}
\end{equation*}

To conclude:

\begin{enumerate}
\item No evidence that space-time symmeries are violated by fundamental laws
of nature. The Translational and Rotational symmetries imply that space is
homogeneous and isotropic. 
\begin{align*}
\text{Translational Symmetry}& \Rightarrow \text{Energy Momentum Conservation%
} \\
\text{Rotational Symmetry}& \Rightarrow \text{Angular Momentum Conservation}
\end{align*}%
If we examine the light emitted by a distant object billions of light years
away, we find that atoms have been following the same laws as they are here
and now. (Translational Symmetry)

\item Discrete Symmetries are not universal; both C and P are violated in
the weak interaction but repsected by electromagnetic and strong
interactions. There is no evidence for violation of time reversal invariance
by any of the fundamental laws of nature.

\item Basic weak interaction Lagrangian is CP conserving. CP violation in
weak interactions is a consequence of mismatch between mass eigenstates and
CP eigenstates and or mismatch between weak and mass eigenstates at quark
level. There is no evidence of CP violation in Lepton sector. There is no
evidence that CP invariance is violated by any of the fundamentals laws of
nature as implied by CPT invariance and T-invariance.

\item CP violation in weak decays is an example where basic laws are CP
invariant but states at quark level contain CP violating phases.

\item The fundamental interaction governing atoms and molecules is the
electromagnetic interaction which does not violate bilateral symmetry
(left-right symmetry). In nature we find organic molecules in asymmetric
form, i.e. left handed or right handed. This is another example where the
basic laws governing these molecules are bilateric symmetric but states are
not. (Asymmetric intial conditions?)

\item \textbf{Baryon Asymmetry of the Universe: Baryogenesis}: No evidence
for existence of antibaryons in the universe. $\eta = n_{B}/n_{\gamma} \sim
3 \times 10^{-10}$. The universe started with a complete matter antimatter
symmetry in big bang picture. In subsequent evolution of the universe, a net
baryon number is generated. This is possible provided the following
conditions of Sakharov are satisfied

\begin{enumerate}
\item There exists a baryon number violating interaction.

\item There exist C and CP violation to induce the asymmetry between
particle and antiparticle processes.

\item Departure from thermal equilibrium of X-particles which mediate the
baryon number violating interactions.
\end{enumerate}

\item There seems to be no connection between CP violation required by
baryogenesis and CP violation observed in weak decays.
\end{enumerate}

\begin{center}
\pagebreak

{\Huge \ Selective List of References.}
\end{center}

\bigskip

\bigskip

\textbf{Figure Captions:}

\begin{enumerate}
\item[Figure 1] The Unitarity triangle

\item[Figure 2] The Box Diagram

\item[Figure 3] The Tree Diagram

\item[Figure 4] The Penguin Diagram

\item[Figure 5] (a) $W$-exchange diagram for $B_{q}^{0}\rightarrow N_{1}\bar{%
N}_{2}\left( M_{f}\right) ;$

(b) $W$-exchange diagram for $B_{q}^{0}\rightarrow N_{1}\bar{N}_{2}\left(
M_{f}^{\prime }\right) \ $

\item[Figure 6] Annihilation diagram for $B^{-}\rightarrow N_{1}\bar{N}_{2}$

\item[Figure 7] Plot of equation $r_{f}\cos \delta _{\left( f\right) }=\cos
\alpha $ for different values of $r.$ For $80^{o}\leq \alpha \leq 103^{o}.\ $%
Where solid curve, dashed curve, dashed doted curve, dashed bouble doted and
double dashed doted curve are corresponding to $r=0.1,\ r=0.15,\ r=0.2,\
r=0.25$ and $r=0.3$ respectively.
\end{enumerate}

\end{document}